\documentclass{aa}
\usepackage{graphicx}
\usepackage{txfonts}
\usepackage[utf8]{inputenc}
\bibpunct{(}{)}{;}{a}{}{,}
\usepackage{graphicx,natbib,url,twoopt,subcaption}
\usepackage[T1]{fontenc}
\usepackage{rotating,booktabs}
\usepackage{float}
\usepackage{adjustbox}
\usepackage{xcolor}
\usepackage{amsmath} 
\usepackage{gensymb}

\begin{document}

   \title{The discovery of two new benchmark brown dwarfs with precise dynamical masses at the stellar-substellar boundary \thanks{Based on observations collected with SPHERE mounted on the VLT at Paranal Observatory (ESO, Chile) under program 0103.C-0199(A) (PI: Rickman), and 105.20SZ.001 (PI: Rickman) as well as observations collected with the CORALIE spectrograph mounted on the 1.2~m Swiss telescope at La Silla Observatory.} \thanks{The radial velocity measurements, reduced images, and additional data products discussed in this paper are available on the DACE web platform at \url{https://dace.unige.ch/.} and the links to individual targets are listed in Appendix~\ref{appendix_a}.}}

   \author{E. L. Rickman \inst{1}, W. Ceva \inst{2}, E. C. Matthews \inst{3, 2}, D. S\'{e}gransan \inst{2}, B. P. Bowler \inst{4}, T. Forveille \inst{5}, K. Franson \inst{4}, J. Hagelberg \inst{2}, S. Udry \inst{2}, A. Vigan \inst{6}}

   \institute{European Space Agency (ESA), ESA Office, Space Telescope Science Institute, 3700 San Martin Drive, 
    Baltimore, MD 21218, USA \email{erickman@stsci.edu}
   \and
   D\'{e}partment d'astronomie de l’Universit\'{e} de Gen\`{e}ve, Chemin Pegasi 51, 1290 Versoix, Switzerland
   \and
   Max-Planck-Institut f\"{u}r Astronomie, Königstuhl 17, D-69117 Heidelberg, Germany
   \and
   Department of Astronomy, The University of Texas at Austin, Austin, TX 78712, USA
   \and
   Univ. Grenoble Alpes, CNRS, IPAG, 38000 Grenoble, France
   \and
   Aix Marseille Univ, CNRS, CNES, LAM, Marseille, France}

   \date{Received; accepted}
   \authorrunning{Rickman et al.}
   \titlerunning{The discovery of two new benchmark brown dwarfs}

 
  \abstract
   {}
   {Measuring dynamical masses of substellar companions is a powerful tool for testing models of mass-luminosity-age relations as well as for determining observational features that constrain the boundary between stellar and substellar companions. In order to dynamically constrain the mass of such companions, we use multiple exoplanet measurement techniques to remove degeneracies in the orbital fits of these objects and place tight constraints on their model-independent masses.}
   {We combined long-period radial velocity data from the CORALIE survey with relative astrometry from direct imaging with VLT/SPHERE as well as with astrometric accelerations from \emph{Hipparcos}-\emph{Gaia} eDR3 to perform a combined orbital fit and measure precise dynamical masses of two newly discovered benchmark brown dwarfs.}
   {We report the discovery of HD~112863~B and HD~206505~B, which are two new benchmark likely brown dwarfs that sit at the substellar-stellar boundary, with precise dynamical masses. We performed an orbital fit that yielded the dynamical masses of $77.1^{+2.9}_{-2.8}~M_{\rm{Jup}}$ and $79.8\pm1.8~M_{\rm{Jup}}$ for HD~112863~B and HD~206505~B, respectively. We determined the orbital period of HD~112863~B to be $21.59\pm0.05$~years and the orbital period of HD~206505~B to be ${50.9}_{-1.5}^{+1.7}$~years. From the $H$ and $K$ band photometry from IRDIS data taken with VLT/SPHERE, we estimate the spectral types of both HD~112863~B and HD~206505~B to be early-mid L-types.}
   {}


   \maketitle
   
%

\section{Introduction}

Companions that have precise model-independent masses and determined ages, known as benchmark brown dwarfs, are fundamental in testing substellar evolutionary models. Such objects are key in placing constraints on the mass-luminosity-age relations of brown dwarfs that are otherwise plagued by a lack of observational constraints \citep{1997ApJ...482..442B,2007ApJ...655..541M,2014MNRAS.437.1378M}.

In order to measure precise dynamical masses of benchmark brown dwarfs, a combination of radial velocity, relative astrometry, and absolute astrometry data can be used to constrain the orbital parameters of such objects and therefore reveal their model-independent masses. Radial velocity measurements provide the minimum mass ($m\sin i$) of an unseen companion around a host star with an unknown orbital inclination ($i$). Proper motions measured from the combination of \emph{Gaia} \citep{2016A&A...595A...2G} and \emph{Hipparcos} \citep{1997A&A...323L..49P} break the degeneracy of the unknown orbital inclination, giving the dynamical mass of the companion. Furthermore, in cases where direct imaging of such companions is possible, we gain additional astrometry relative to their host star that can tightly constrain the orbital parameters and therefore precise mass measurements. Relative astrometry provides not only additional constraints on the orbit of the system and the dynamical mass of the companion, but it also provides photometry, revealing an estimate of the spectral type of a detected companion.

Directly detecting such substellar companions, however, does not come without its challenges. Previously, many direct imaging searches have adopted a "blind" survey approach, but the detection rate for such an approach has been low \citep[e.g.,][]{2018haex.bookE.155B}. To unveil the substellar companion population, an approach of target selection using precursor measurements is fundamental in increasing the efficiency of direct imaging of substellar companions.

In this work, we assess the feasibility of directly imaging companions that show indirect detection with radial velocities and absolute astrometry by performing an orbital fit. The predicted relative separation and estimate of the dynamical mass are assessed against the expected contrast for VLT/SPHERE, as demonstrated in \cite{2022A&A...668A.140R}, and we take coronagraphic imaging observations to confirm the detection of these objects directly.

As a result of adopting this methodology, we present in this paper the direct detection of two new benchmark brown dwarfs, HD~112863~B and HD~206505~B, and we give their dynamical masses. These targets join the short but increasing list of substellar objects with known dynamical masses \citep[e.g.][]{2018A&A...614A..16C,2018AJ....155..159B,2019AJ....158..140B,2020A&A...639A..47M,2020A&A...635A.203R,2021AJ....162..301B,2022MNRAS.513.5588B,2022AJ....163...50F,2023AJ....165...39F}. The two brown dwarfs that we present in this paper sit right at the boundary of the hydrogen-burning limit of $\sim75-80~M_{\rm{Jup}}$ \citep{2001RvMP...73..719B,2008ApJ...689.1327S,2015A&A...577A..42B,2017ApJS..231...15D,2019ApJ...879...94F}, making them key objects to study the boundary between what is considered a brown dwarf and what is considered a very low-mass star. Additionally, these objects can be used to empirically validate mass-luminosity-age relations of substellar objects, a crucial step in understanding evolutionary models of brown dwarfs more broadly.

The paper is organized as follows. The properties of the host stars are outlined in Section~\ref{sec:stellar_params}. In Section~\ref{sec:obs}, we present the radial velocities, astrometry, and direct imaging observations and data reduction. In Sections~\ref{sec:orbital parameters}, we present the detections and orbital solutions of HD~112863~B and HD~206505~B. A brief discussion on the implications of our findings and the conclusions of this paper are presented in Section~\ref{sec:conclusion}.

\section{Characteristics of stellar hosts} \label{sec:stellar_params}

The spectral types and the color indices of the primary stars were obtained from the \emph{Hipparcos} catalog \citep{1997A&A...323L..49P}. The $V_T$ band magnitudes were taken from the Tycho-2 catalog \citep{2000A&A...355L..27H}. The luminosities $L$ and effective temperatures ($T_{\mathrm{eff}}$) for the two host stars are from the \emph{Gaia} data release 2 \citep[DR2;][]{2018A&A...616A...1G}, while the astrometric parallaxes ($\pi$) are from the \emph{Gaia} early data release 3 \cite[eDR3;][]{2021A&A...649A...1G}.

The $v\sin(i)$ of HD~112863~A and HD~206505~A were calculated through the calibration of the width of the cross-correlation function (CCF) of the CORALIE spectrograph as described in \cite{2001A&A...373.1019S} and \cite{marmier_phd_thesis}. The stellar surface gravities ($\log g$) and metallicity ([Fe/H]) values are taken from \citet{2014A&A...566A..83M}.

The ages and masses of the two primary stars were determined using the Geneva stellar isochrones \citep{2012A&A...537A.146E,2013A&A...558A.103G}, which utilizes a Markov chain Monte Carlo (MCMC) approach.\footnote{The tools using this approach can be found here: \url{https://www.unige.ch/sciences/astro/evolution/en/database/syclist/}.} We ran the MCMC with a chain length of 100,000 in both cases and with the Gaussian priors input for the [Fe/H], $T_{\rm{eff}}$, and $V$-band magnitude as listed in Table~\ref{tab:stellar_params}. The resulting values for the mass, radius, and age of the primary stars are shown in Table~\ref{tab:stellar_params}.

\begin{table}[]
    \centering
    \caption{Observed and inferred stellar parameters for host stars HD~112863~A and HD~206505~A.}
    \begin{tabular}{cccc}
    \hline
    \hline
     Parameters & Units & HD~112863 & HD~206505 \\
     \hline
     Sp. Type \tablefoottext{a} & & K1V & K0V \\
     $V_T$ \tablefoottext{b} & & 8.78 & 8.82 \\
     $B-V$ \tablefoottext{c} & & 0.779 & 0.819 \\
     $\alpha$ \tablefoottext{d} & J2000 & 12 59 45.51 & 21 48 30.98 \\
     $\delta$ \tablefoottext{d} & J2000 & -04 25 49.05 & -78 25 59.68 \\
     $\mu_{\alpha} \cos \delta$ \tablefoottext{d} & $[\rm{mas~yr}^{-1}]$ & 15.968 & -133.702 \\
     $\mu_{\delta}$ \tablefoottext{d} & $[\rm{mas~yr}^{-1}]$ & -28.770 & 119.298 \\
     $U$ & [$\rm{km~s}^{-1}$] & 1.49 \tablefoottext{e} & 36.0 \tablefoottext{f} \\
     $V$ & [$\rm{km~s}^{-1}$] & 2.10 \tablefoottext{e} & 18.0 \tablefoottext{f} \\
     $W$ & [$\rm{km~s}^{-1}$] & -11.68 \tablefoottext{e} & 0.0 \tablefoottext{f} \\
     $\pi$ \tablefoottext{g} & [mas] & $26.96\pm0.03$ & $22.77\pm0.02$ \\
     $L$ \tablefoottext{h} & [$\rm{L}_{\odot}$] & $0.429\pm0.001$ & $0.595\pm0.001$ \\
     $T_{\rm{eff}}$ \tablefoottext{h} & [K] & $5342^{+76}_{-43}$ & $5377^{+50}_{-58}$ \\
     $\log g$ \tablefoottext{h} & [cgs] & $4.57\pm0.07$ & $4.46\pm0.07$ \\
     $[\rm{Fe/H}]$ \tablefoottext{i} & [dex] & $-0.11\pm0.03$ & $0.11\pm0.03$ \\
     $v \sin i$ \tablefoottext{j} & [$\rm{km s^{-1}}$] & 3.73 & 2.01 \\
     \hline
     $M_{*, \rm{isochronal}}$ & [$\rm{M}_{\odot}$] & $0.85\pm0.02$ & $0.93\pm0.02$ \\
     $R_{*, \rm{isochronal}}$ & [$\rm{R}_{\odot}$] & $0.76\pm0.02$ & $0.88\pm0.02$ \\
     Age$_{\rm{isochronal}}$ & [Gyr] & $3.31\pm2.91$ & $3.94\pm2.51$ \\
    \hline
    \end{tabular}
    \tablefoot{
    \tablefoottext{a}{Parameters taken from \citet{1975mcts.book.....H} and \citet{1999MSS...C05....0H}.}
    \tablefoottext{b}{Parameters taken from the Tycho-2 catalog \citep{2000A&A...355L..27H}.}
    \tablefoottext{c}{Parameters taken from the \emph{Hipparcos} catalog \citep{1997A&A...323L..49P}.}
    \tablefoottext{d}{Parameters taken from \citet{2020yCat.1350....0G}.}
    \tablefoottext{e}{Parameters taken from \citet{2013yCat..35520027M}.}
    \tablefoottext{f}{Parameters taken from \citet{2009yCat.5130....0H}.}
    \tablefoottext{g}{Parameters taken from \emph{Gaia} early data release 3 \citep{2021A&A...649A...1G}.}
    \tablefoottext{h}{Parameters taken from \emph{Gaia} data release 2 \citep{2018A&A...616A...1G}.}
    \tablefoottext{i}{Parameters taken from \citet{2014A&A...566A..83M}.}
    \tablefoottext{j}{Parameters derived using CORALIE CCF.}}
    \label{tab:stellar_params}
\end{table}

\section{Observations and data reduction} \label{sec:obs}

\begin{figure*}
    \centering
    \begin{subfigure}[t]{0.24\linewidth}
    \includegraphics[width=\textwidth]{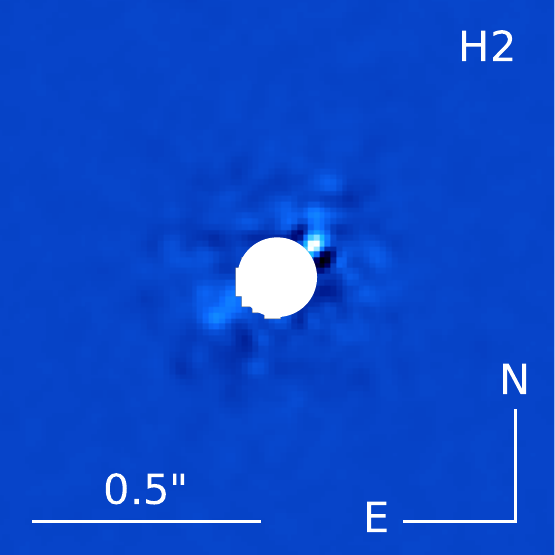}
    \caption{2021-04-07}
    \end{subfigure}
    \begin{subfigure}[t]{0.24\linewidth}
    \includegraphics[width=\textwidth]{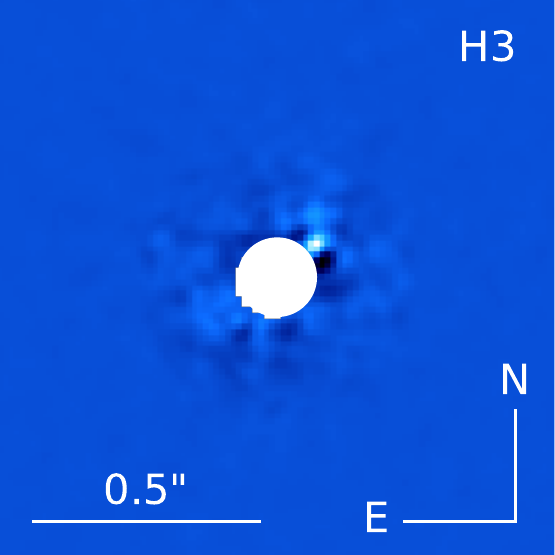}
    \caption{2021-04-07}
    \end{subfigure}
    \begin{subfigure}[t]{0.24\linewidth}
    \includegraphics[width=\textwidth]{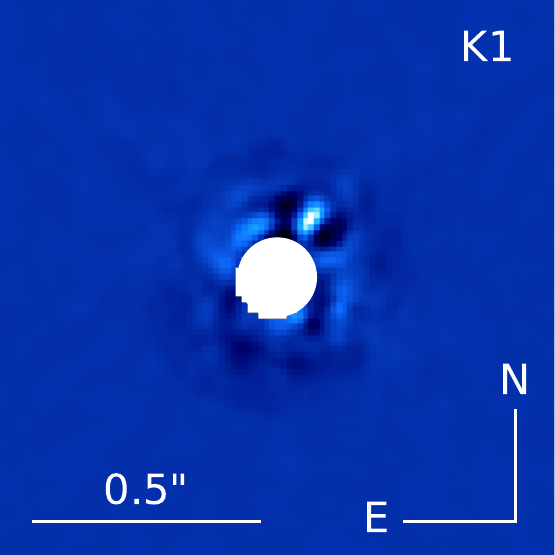}
    \caption{2022-01-30}
    \end{subfigure}
    \begin{subfigure}[t]{0.24\linewidth}
    \includegraphics[width=\textwidth]{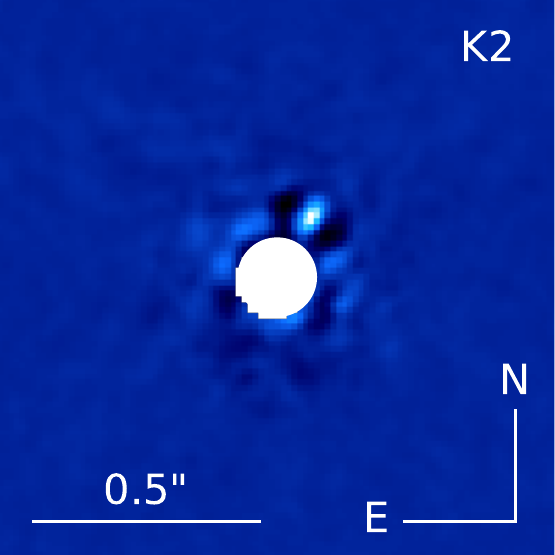}
    \caption{2022-01-30}
    \end{subfigure}
    \begin{subfigure}[t]{0.24\linewidth}
    \includegraphics[width=\textwidth]{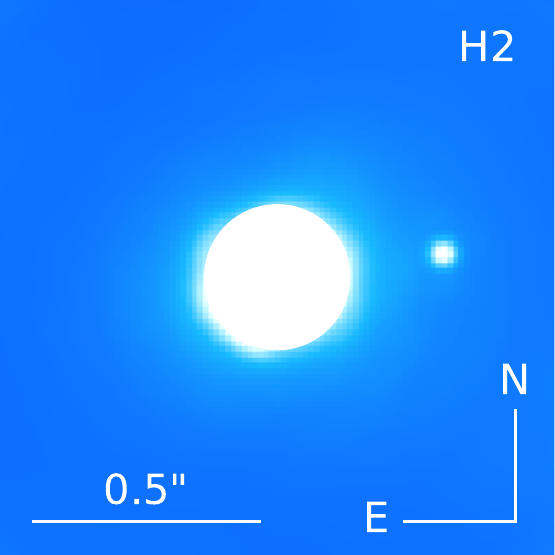}
    \caption{2019-08-06}
    \end{subfigure}
    \begin{subfigure}[t]{0.24\linewidth}
    \includegraphics[width=\textwidth]{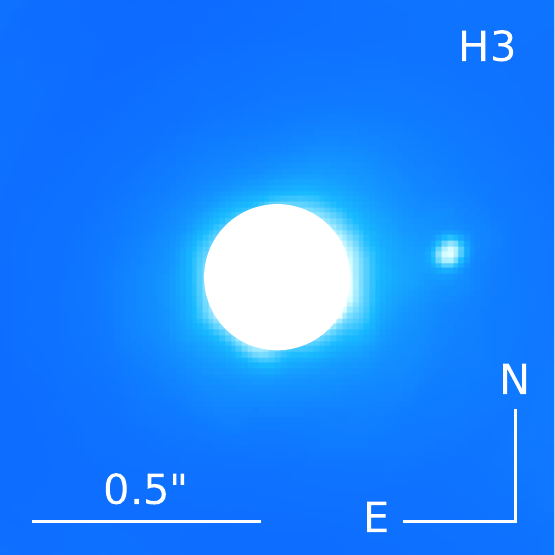}
    \caption{2019-08-06}
    \end{subfigure}
    \begin{subfigure}[t]{0.24\linewidth}
    \includegraphics[width=\textwidth]{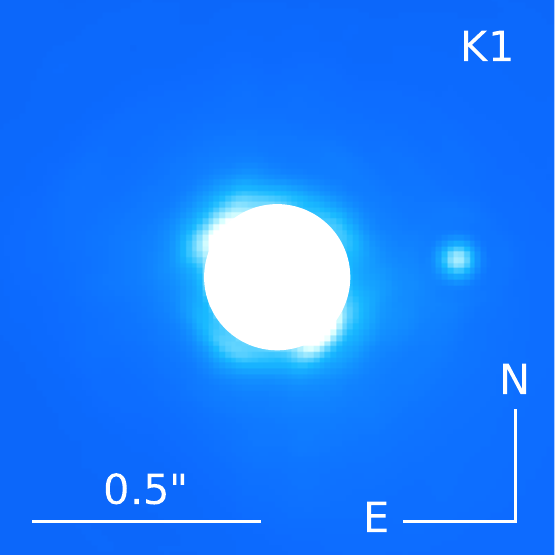}
    \caption{2021-07-01}
    \end{subfigure}
    \begin{subfigure}[t]{0.24\linewidth}
    \includegraphics[width=\textwidth]{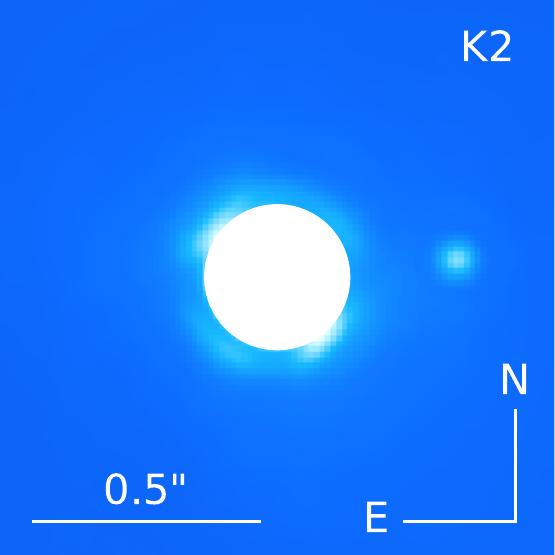}
    \caption{2021-07-01}
    \end{subfigure}
    \caption{High-contrast images of HD~112863~B and HD~206505~B taken with VLT/SPHERE IRDIS coronagraphy. The date of each image is shown in each sub-caption in the format YYYY-MM-DD. The filter used is shown in each image. The primary star in each image is masked behind the white circle where the coronagraph is. \textbf{Top:} VLT/SPHERE images of HD~112863~B. \textbf{Bottom:} VLT/SPHERE images of HD~206505~B.}
    \label{fig:images}
\end{figure*}

\begin{table*}
    \centering
        \caption{Relative astrometry and photometry of HD~112863~B and HD~206505~B. We note that the values listed for HD~112863~B were obtained after correcting for the attenuation by the coronagraph that occurs within the IWA of $150$~mas.}  
    \begin{tabular}{ccccccc}
    \hline
    \hline
        Companion & Date (yyyy-mm-dd) & Filter & $\rho$ (mas) & $\theta$ (deg) & Contrast & Abs. Mag \\
    \hline
        HD~112863~B & 2021-04-07 & H2 & $105.3\pm3.2$ & $309.80\pm1.67$ & $7.65\pm0.05$ & $11.69\pm0.07$ \\
        HD~112863~B & 2021-04-07 & H3 & $106.6\pm3.1$ & $309.27\pm1.65$ & $7.454\pm0.03$ & $11.49\pm0.06$ \\
        HD~112863~B & 2022-01-30 & K1 & $148.1\pm3.3$ & $329.44\pm1.21$ & $6.90\pm0.04$ & $10.84\pm0.04$ \\
        HD~112863~B & 2022-01-30 & K2 & $144.5\pm3.5$ & $327.40\pm1.24$ & $6.64\pm0.05$ & $10.58\pm0.05$ \\
        HD~206505~B & 2019-08-06 & H2 & $376.6\pm3.1$ & $276.84\pm0.48$ & $7.57\pm0.003$ & $11.31\pm0.05$ \\ 
        HD~206505~B & 2019-08-06 & H3 & $376.4\pm3.1$ & $276.19\pm0.48$ & $7.43\pm0.003$ & $11.17\pm0.05$ \\ 
        HD~206505~B & 2021-07-01 & K1 & $404.3\pm3.1$ & $274.24\pm0.45$ & $7.20\pm0.003$ & $10.85\pm0.03$ \\ 
        HD~206505~B & 2021-07-01 & K2 & $403.1\pm3.1$ & $274.27\pm0.45$ & $6.96\pm0.003$ & $10.61\pm0.03$ \\ 
        \hline
    \end{tabular}
    \label{tab:astro_photo}
\end{table*}

\subsection{Radial velocities}

We used radial velocity measurements taken from the CORALIE survey \citep{2000fepc.conf..548Q,2000fepc.conf..571U}, which is an ongoing radial velocity survey with a wealth of data taken in the southern hemisphere since June 1998. The survey utilizes the CORALIE spectrograph on the Swiss/Euler 1.2~m telescope at La Silla Observatory in Chile. The radial velocity survey includes a sample of 1647 main-sequence stars within 50~pc of the Sun.

The CORALIE spectrograph underwent two major upgrades, one in June 2007 \citep{2010A&A...511A..45S} and the second in November 2014, to improve its overall performance. These upgrades introduced small offsets in the measured radial velocities. Due to this, we treated radial velocity data from the CORALIE spectrograph as three separate instruments, referring to the original CORALIE spectrograph, the 2007 upgrade, and the 2014 upgrade as CORALIE-98 (C98), CORALIE-07 (C07), and CORALIE-14 (C14), respectively. All the data products presented in this paper are available at the Data and Analysis Center for Exoplanets (DACE).\footnote{The data are available at the DACE, which can be accessed at: \url{https://dace.unige.ch}, with the individual targets available for HD~112863 and HD~206505 at \url{https://dace.unige.ch/radialVelocities/?pattern=HD112863} and \url{https://dace.unige.ch/radialVelocities/?pattern=HD206505}, respectively.}

The radial velocity data were reduced using the CORALIE automated pipeline \citep{2000SPIE.4009...61W}. This pipeline measures the CCF, the full width at half maximum (FWHM), the bisector, and the $H_{\alpha}$ chromospheric activity indicator. We used these indicators to ensure that any observed periodic signals are not due to any stellar activity of the host star, which could mimic the expected radial velocity signal of an unseen companion. This is described in more detail in Appendix~\ref{appendix_RV}. We also checked for the presence of any additional planetary signals at shorter orbital periods in the radial velocity data, and we did not find evidence of any additional companions in either system, as shown in Appendix~\ref{appendix_RV}. From the over 20-year baseline of CORALIE radial velocity data, we selected candidates that show signs of hosting long-period companion candidates through either linear or quadratic trends, as shown previously in \cite{2019A&A...625A..71R}, that could potentially be directly detected with high-contrast imaging.

\subsection{Absolute astrometry} \label{sec:abs_astrometry}

In order to utilize astrometric acceleration information from \emph{Hipparcos} and \emph{Gaia}, we used the \emph{Hipparcos}-\emph{Gaia} catalog of accelerations \citep[HGCA;][]{2021ApJS..254...42B}. The HGCA is a cross-calibration of \emph{Hipparcos} \citep{1997ESASP1200.....E,2007A&A...474..653V} and Gaia eDR3 \citep{2016A&A...595A...1G,2021A&A...649A...1G,2021A&A...649A...2L} that places both on a common reference frame with calibrated uncertainties. Each star in the HGCA has three proper motions: a \emph{Hipparcos} proper motion near 1991.25, a Gaia proper motion near 2016.0, and an average proper motion between both epochs, calculated as the positional difference between \emph{Hipparcos} and Gaia scaled by the time baseline. The long baseline between these missions enabled us to find astrometric accelerators that could be hosting companion candidates and that are able to be directly detected, much like in the case of long baseline radial velocity detections.

\subsection{Direct imaging} \label{sec:imaging}

We elected to observe targets that were predicted to be directly detectable from their radial velocity and absolute astrometric measurements. Incorporating the astrometric accelerations from the HGCA into the radial velocity information enables a more comprehensive characterization of the orbital parameters of the stellar companion. The sensitivity of the combination of \textit{Hipparcos} and \textit{Gaia} proper motions to orbital periods has been demonstrated for orbital periods extending to several hundreds of years \citep{2018ApJS..239...31B,2021ApJS..254...42B}. This means that the expected position of the companion relative to the host star can be well predicted before the direct detection itself, as described in Appendix~\ref{appendix_astrometric_predictions}. 

Knowing the predicted position and therefore the relative angular separation of a companion relative to its host star allows the feasibility of the direct detection to be assessed against the known coronagraphic inner working angle (IWA) as well as the expected contrast ratio against measured contrast curves, as demonstrated in \citep{2022A&A...668A.140R}. Based on these criteria, we observed HD~112863 and HD~206505 with VLT/SPHERE \citep{2019A&A...631A.155B} via the extreme adaptive optics system at the VLT under programs 0103.C-0199(A) (PI: Rickman) and 105.20SZ.001 (PI: Rickman).

We used the VLT/SPHERE dual-band imaging mode \citep{2010MNRAS.407...71V}, which makes use of the InfraRed Dual-Band Imager and Spectrograph \citep[IRDIS;][]{2008SPIE.7014E..3LD} in the $H2$ and $H3$ bands ($\lambda_{H2}=1.593~\mu$m, $\lambda_{H3}=1.667~\mu$m), as well as the $K1$ and $K2$ bands ($\lambda_{K1}=2.110~\mu$m, $\lambda_{K2}=2.251~\mu$m). In this paper, we focus on the VLT/SPHERE IRDIS data. There are additional VLT/SPHERE integral field spectrograph \citep[IFS;][]{2008SPIE.7014E..3EC} data that will be presented with a more in-depth analysis of the spectroscopic components in Ceva et al. (in prep.).

The data were reduced using the Geneva Reduction and Analysis Pipeline for High-contrast Imaging of planetary Companions \citep[GRAPHIC;][]{2016MNRAS.455.2178H}. GRAPHIC performs sky subtraction, flat fielding, bad pixel cleaning, and anamorphic distortion correction \citep{2016SPIE.9908E..34M}. We then used principal component analysis \citep[PCA;][]{2012ApJ...755L..28S, 2012MNRAS.427..948A} and angular differential imaging \citep[ADI;][]{2006ApJ...641..556M} on the reduced data to remove point spread function (PSF) residuals. The detection images of both companions for both bands in both epochs are shown in Fig.~\ref{fig:images}. 




The relative astrometry and photometry of the companions were calculated using the negative fake planet injection technique as used in \cite{2011A&A...528L..15B}, and in particular, we followed the same procedure as used in \cite{2020A&A...635A.203R}.  The forward models of the PSFs were generated using observations of the target stars through a neutral density filter while not behind the coronagraph and with shorter exposure times than the standard science observations. We therefore scaled the stellar PSFs to correct for these differences in exposure time and filter transmission prior to insertion into the science images.\footnote{\label{SPHfilt1}The corrections for the Neutral Density (ND) filter transmission make use of the filter curves available at \url{https://www.eso.org/sci/facilities/paranal/instruments/sphere/inst/filters.html}.}  Given the large amount of observations of HD~206505 and the brightness of HD~206505~B, the reduced frames of the HD~206505 were cropped and binned prior to PSF insertion in order to reduce computation time while still obtaining precise relative astrometry and photometry.  

Since HD~112863~B lies within the $150$~mas IWA of SPHERE's $H23$ and $K12$ dual-band imaging modes\footnote{From the VLT SPHERE User Manual, 16th release} in both epochs of observations (see projected angular separation values ($\rho$) in Table~\ref{tab:astro_photo}), its flux is attenuated by the coronagraph. As the transmissivity of the coronagraph changes on scales smaller than a PSF, the native companion PSF is therefore distorted, and both astrometric and photometric measurements of the companion are biased.  To correct for this effect, we used the SPHERE $H23$ and $K12$ coronagraphic transmission profiles (Vigan 2023, private communication) to create radial coronagraphic transmission images and then divided the reduced frames of HD~112863 by these radial coronagraphic transmission images. We then performed the same process as described above of injecting scaled stellar PSF images into the coronagraphic transmission-corrected reduced frames. The effect of this was much less acute for the second HD~112863 epoch in the $K12$ band, as the separation is greater than the first epoch in the $H23$ band. Our coronagraph transmission-correction approach is discussed in more detail in Ceva et al. (in prep).

The separation and position angle detector positions were then converted into on-sky separation and position angles by accounting for the plate scale of each band, the anamorphic distortion, the true north offset, and the pupil offset using the values found in \cite{2016SPIE.9908E..34M}.  Additionally, a systematic uncertainty of $\pm3$~mas for the positions of the target stars was folded into the positional errors of the companions \citep{2016A&A...587A..55V}. The resulting astrometry and photometry of both companions for both bands in both epochs are shown in Table~\ref{tab:astro_photo}.  We note that the uncertainty on the separations for nearly all of the measurements is similar ($\sim3~\textrm{mas}$). This is because the errors of the separation values prior to conversion to on-sky values are on the scale of $\sim0.1-0.5$~mas such that the $\pm3$~mas uncertainty of the positions of the target stars become the dominant term in the final on-sky separation error of the companions.

\section{Orbital solutions} \label{sec:orbital parameters}

To obtain orbital fits of our observed systems, we used the orbit-fitting code \texttt{orvara} \citep{2021AJ....162..186B}, which has the capability of combining radial velocity data with relative astrometry from direct imaging and absolute astrometry from \emph{Hipparcos} and \emph{Gaia} using a comprehensive MCMC approach.\footnote{\texttt{orvara} can be accessed via GitHub here: \url{https://github.com/t-brandt/orvara}.} In this section, we describe the fitted orbital solutions to HD~112863 and HD~206505. The output of the orbital parameters determined by fitting for the radial velocities, the astrometric accelerations from HGCA, and the relative astrometry from VLT/SPHERE imaging are shown in Table~\ref{tab:orbital_parameters}.

\subsection{HD~112863 (HIP~63419)} \label{sec:HD112863}

\begin{figure*}
    \centering
    \includegraphics[width=0.42\textwidth]{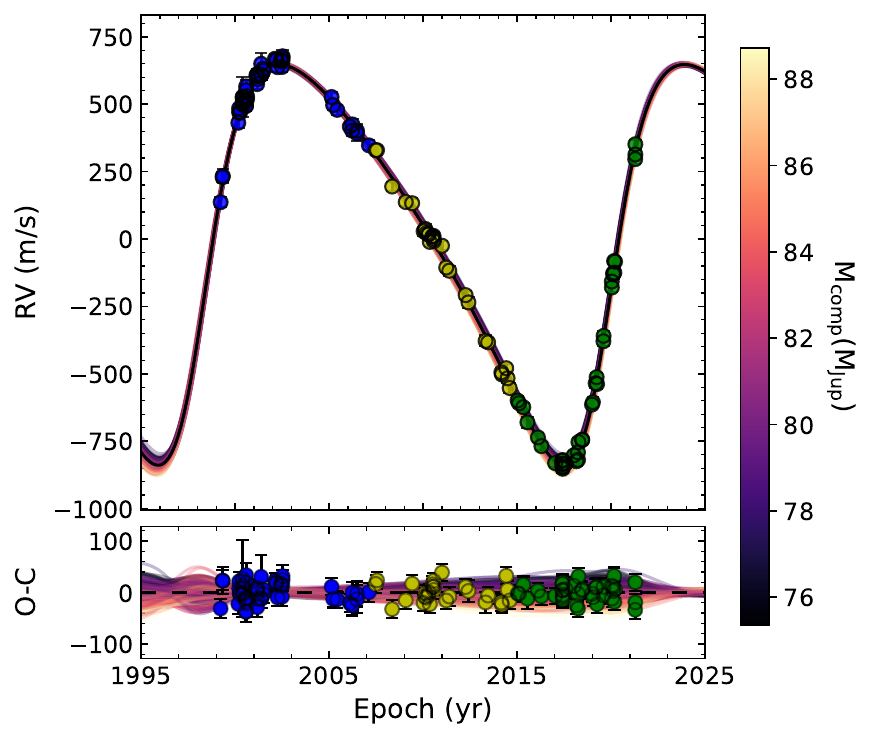}
    \includegraphics[width=0.45\textwidth]{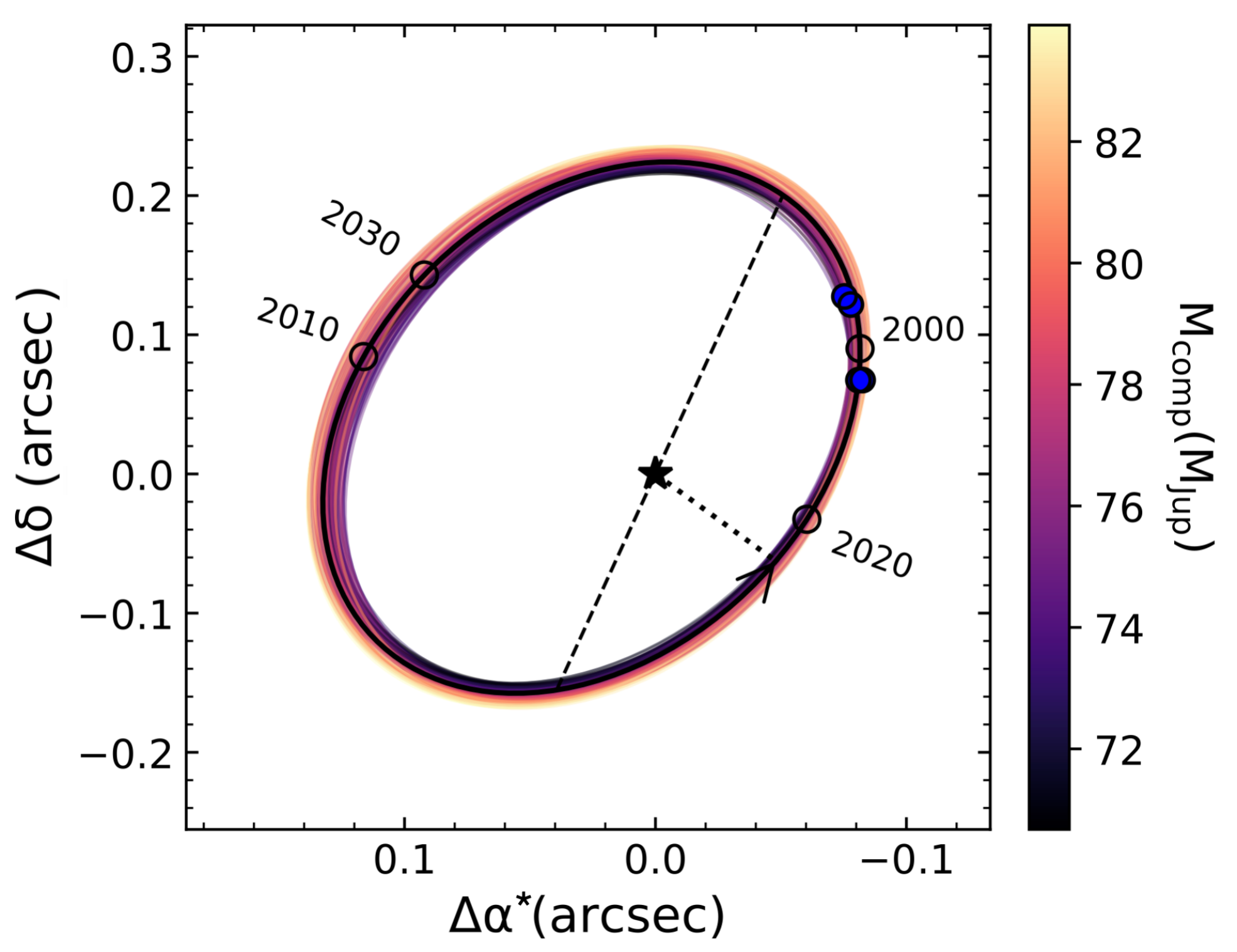}
    \includegraphics[width=0.8\textwidth]{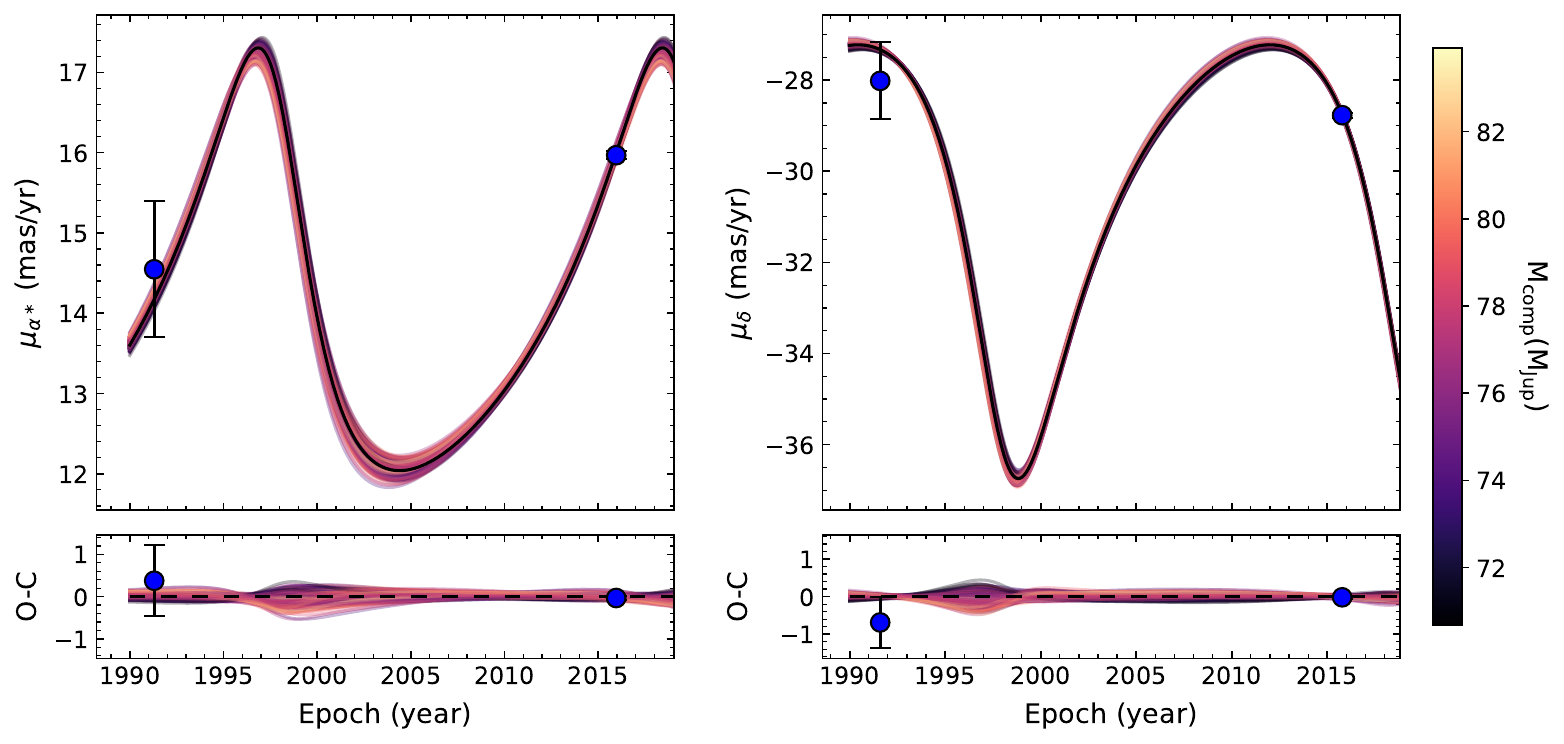}
    \caption{Orbit fits of HD~112863 using the orbit-fitting code \texttt{orvara}. \textbf{Top left:} Radial velocity orbit induced by HD~112863~B over a full orbital period. Shown are the radial velocity data of COR-98 (blue points), COR-07 (yellow points), and COR-14 (green points). The thick line shows the highest likelihood fit; the thin colored lines show 500 orbits drawn randomly from the posterior distribution. \textbf{Top right:} Relative astrometric orbit of HD~112863~B relative to its host star in right ascension ($\Delta \alpha^{*} = \Delta \alpha \cos \delta$) and declination ($\Delta \delta$). The thick black line represents the highest likelihood orbit; the thin colored lines represent 500 orbits drawn randomly from the posterior distribution. Dark purple corresponds to a low companion mass, and light yellow corresponds to a high companion mass. The dotted black line shows the , and the arrow at the periastron passage shows the direction of the orbit. The dashed line indicates the line of nodes. Predicted past and future relative astrometric points are shown by black circles with their respective years, while the observed relative astrometric point from VLT/SPHERE data is shown by the blue-filled data point, where the measurement error is smaller than the plotted symbol. \textbf{Bottom:} Acceleration induced by the companion on the host star as measured from absolute astrometry from \emph{Hipparcos} and \emph{Gaia}. The thick black line represents the highest likelihood orbit; the thin colored lines are 500 orbits drawn randomly from the posterior distribution. The residuals of the proper motions are shown in the bottom panels.}
    \label{fig:HD112863_orbit}
\end{figure*}

The star HD~112863 was monitored with the CORALIE radial velocity survey between March 1999 and June 2023, covering 24 years of observations with 132 radial velocity measurements in total and providing a significant orbital phase coverage of the radial velocities, as shown in Fig.~\ref{fig:HD112863_orbit}. The RVs and astrometric acceleration joint analysis were previously presented in \citet{2023arXiv230316717B}. Here, we present the first direct detection of HD~112863~B and updated orbital parameters that incorporate new relative astrometry from these observations along with the RVs and HGCA data as presented in \citet{2023arXiv230316717B}.

\begin{figure*}
    \centering
    \includegraphics[width=0.40\textwidth]{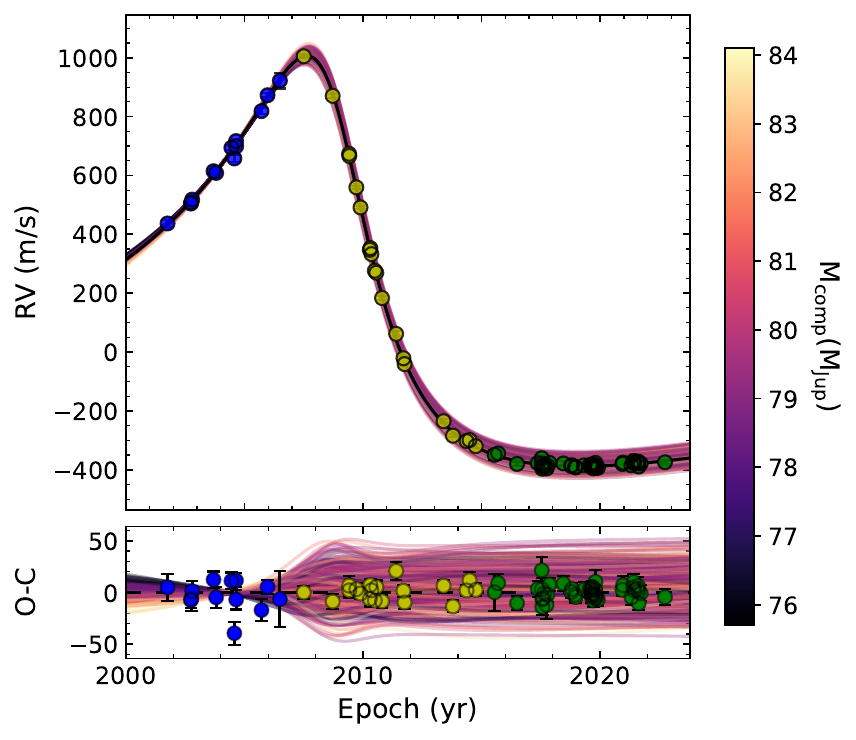}
    \includegraphics[width=0.45\textwidth]{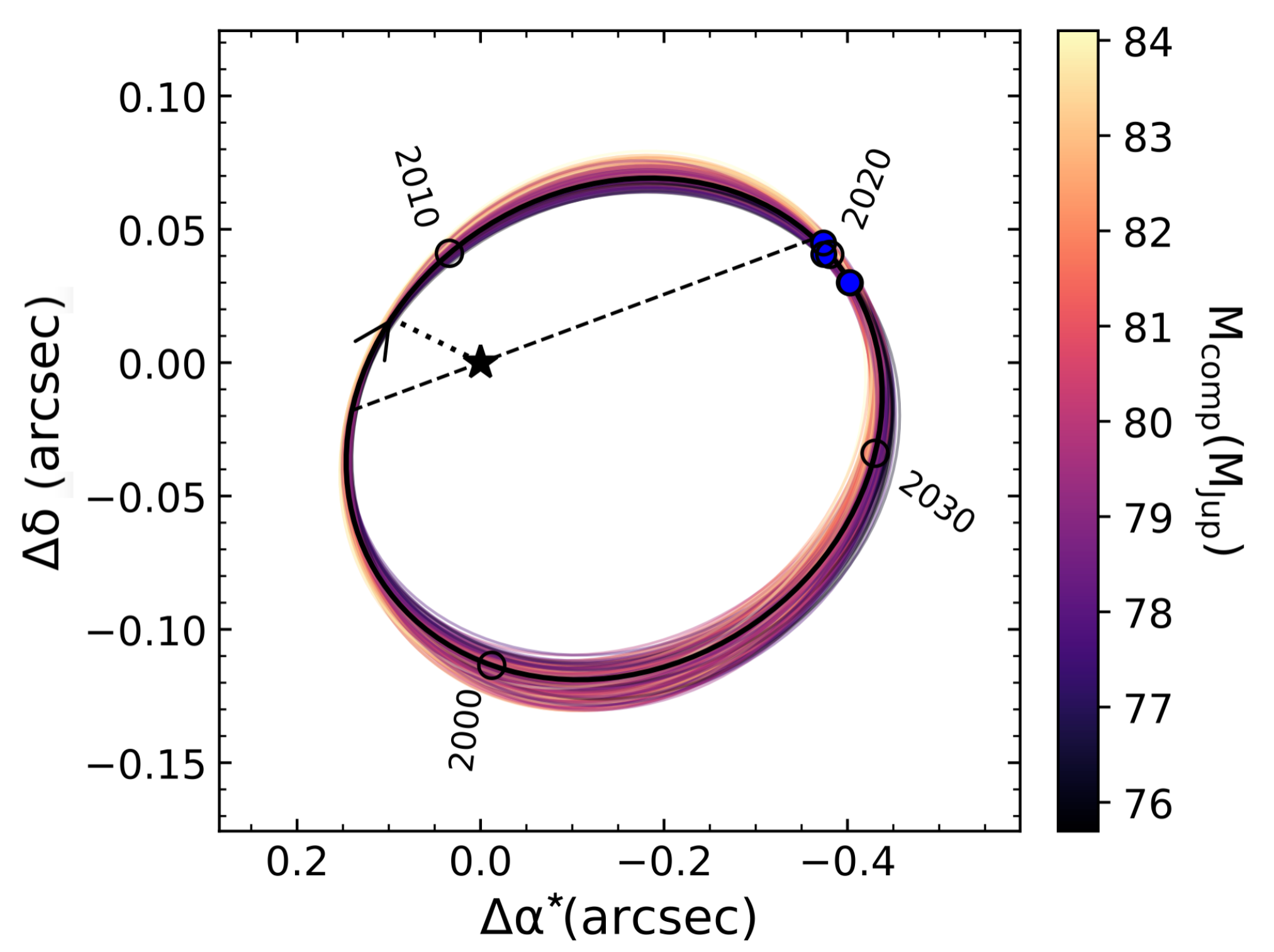}
    \includegraphics[width=0.8\textwidth]{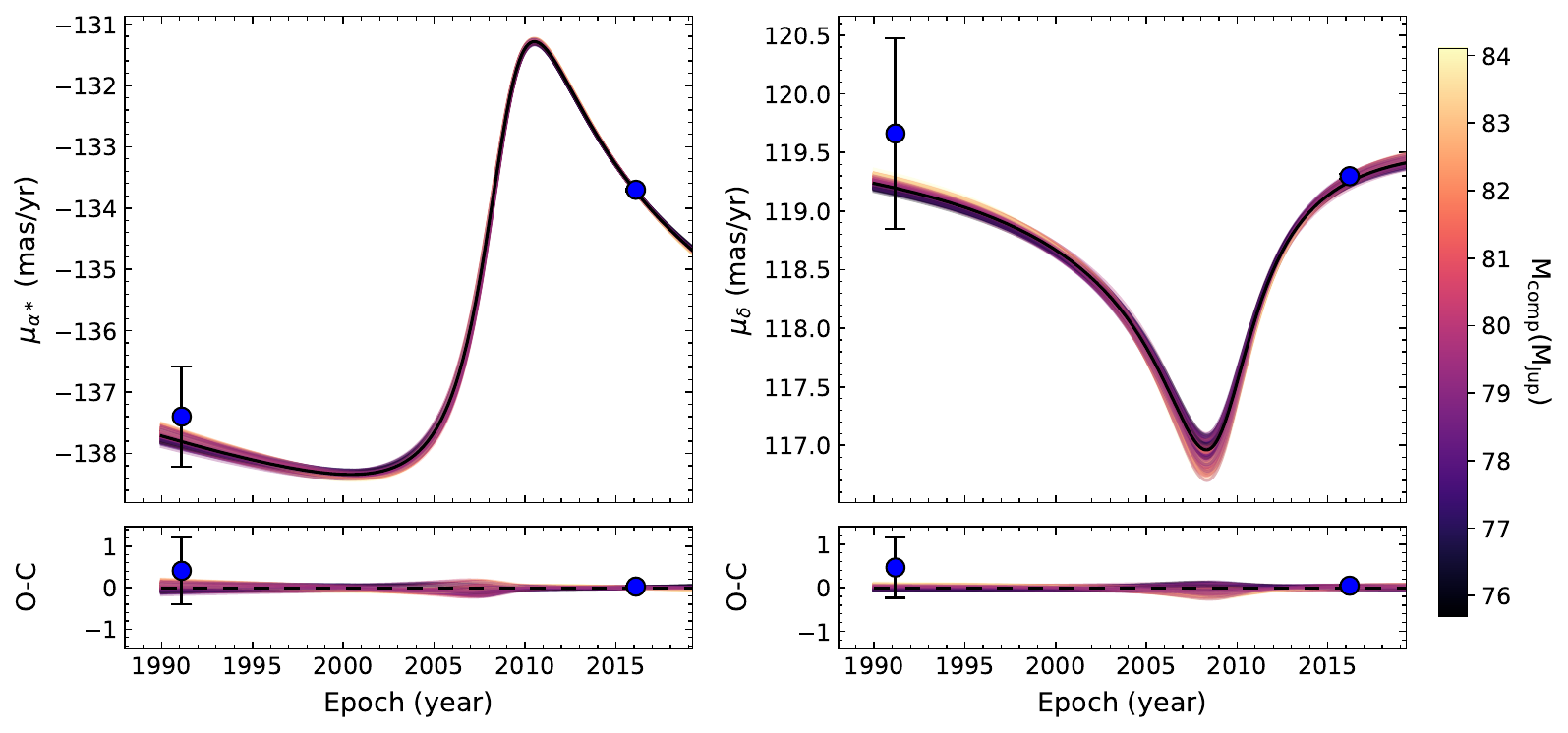}
    \caption{Same as Fig.~\ref{fig:HD112863_orbit} but for HD~206505.}
    \label{fig:HD206505_orbit}
\end{figure*}

We directly imaged HD~112863~B with VLT/SPHERE on 2021-04-07 with IRDIS in $H2$ and $H3$ bands as part of program 105.20SZ.001 (PI: Rickman), as shown in Fig.~\ref{fig:images}, with a total integration time of 4096~seconds. The detection of this object is right at the limit of the IWA of the coronagraph of SPHERE. We measured a projected angular separation of $105.3\pm3.2~$mas and $106.6\pm3.1$~mas for the $H2$ and $H3$ bands, respectively, as outlined in Table~\ref{tab:astro_photo}, which to date is the smallest separation companion directly imaged with SPHERE/IRDIS coronagraphy. We incorporated the coronagraphic transmission correction (see Section~\ref{sec:imaging}) when calculating the astrometry in order to ensure that our values are not biased by any PSF distortion due to the coronagraph. We conducted further follow-up imaging of HD~112863~B on 2022-01-30 in the $K1$ and $K2$ bands as part of program 105.20SZ.001 (PI: Rickman), also shown in Fig.~\ref{fig:images}, with a total integration time of 4864~seconds. This provided additional relative astrometry that could further constrain the orbital parameters and therefore the dynamical mass of the brown dwarf companion as well as extend the photometric baseline, which improves the determination of the spectral type. This is discussed in Section~\ref{sec:CMD}. 

We performed an orbit fit that combined absolute astrometry from \emph{Gaia} \citep{2016A&A...595A...1G} and \emph{Hipparcos} \citep{1997ESASP1200.....E} that made use of the HGCA as described in Section~\ref{sec:abs_astrometry} along with the CORALIE radial velocities \citep{2023arXiv230316717B} and the relative astrometry measured through direct imaging, as shown in Table~\ref{tab:astro_photo}. For the fit, we used the \texttt{orvara} code, employing a parallel-tempered MCMC with 15 temperatures. For each temperature, we used 100 walkers with 40,000 steps per walker thinned by a factor of 50. We used a log-flat prior on the host star mass in order to also measure the mass dynamically.

As the orbit phase is well sampled by the radial velocity measurements, we were able to constrain an orbital period of $21.59\pm0.05$~years, which is in agreement with the orbital period of $21.61\pm0.04$~years from \cite{2023arXiv230316717B}. We measured the dynamical mass of the primary to be $M_{\rm{host}}=0.89^{+0.05}_{-0.04}~M_{\odot}$, which is in agreement with the  from \cite{2023arXiv230316717B} of $0.81\pm0.05~M_{\odot}$, which was derived using the stellar spectral energy distribution (SED). The dynamical mass of the primary star is also in agreement with the isochronal mass measured in this paper of $0.85\pm0.02~M_{\odot}$. The dynamical mass measurement of the BD companion HD~112863~B is $M_{\rm{comp}}=77.1^{+2.9}_{-2.8}~M_{\rm{Jup}}$, which is also in agreement with the RV and astrometrically derived mass from \cite{2023arXiv230316717B} of $73.10\pm3.20~M_{\rm{Jup}}$. Unlike the analysis in this paper, \cite{2023arXiv230316717B} imposed a Gaussian prior on the primary stellar mass for the orbital fit. Despite this, the additional relative astrometric data from imaging adds constraints to the orbital fit that give a dynamical mass of the companion at a marginally higher level of precision than reported in \cite{2023arXiv230316717B} without relying on a constrained prior of the primary star, meaning that we calculated the dynamical mass of the primary as well.

The resulting orbital fits are shown in Fig.~\ref{fig:HD112863_orbit}, and the full orbital parameters are listed in Table~\ref{tab:orbital_parameters}, with the posteriors from the MCMC shown in Fig.~\ref{fig:HD112863_corner}. The relative astrometry in terms of projected angular separation and position angle are shown in Fig.~\ref{fig:HD112863_relsep_pa}.

\begin{table*}
    \caption{Markov chain Monte Carlo orbital posteriors for the orbital fits of each system using \texttt{orvara} \citep{2021AJ....162..186B}.}
    \centering
    \begin{tabular}{ccccc}
    \hline\hline
         Parameters & Units & Prior & HD~112863 & HD~206505 \\
         \hline
         & & Fitted Parameters \\
         \hline
         Companion mass $M_{\rm{comp}}$ & $M_{\rm{Jup}}$ & $1/M_{\rm{comp}}$ (log-flat) & $77.1^{+2.9}_{-2.8}$ & $79.8\pm1.8$ \\
         Host star mass $M_{\rm{host}}$ & $M_{\odot}$ &  $1/M_{\rm{host}}$ (log-flat) & $0.89_{-0.04}^{+0.05}$ & $0.97\pm0.03$ \\
         Parallax $\pi$ & mas & $1/\pi$ (log-flat) & $26.955\pm0.003$ & $22.768\pm0.003$ \\
         Inclination $i$ & $\degree$ & $\mathcal{U}(0,180)$ & $60.46\pm0.64$ & $110.04\pm0.74$ \\
         Semimajor axis $a$ & AU & $1/a$ (log-flat) & $7.66\pm0.13$ & ${13.94}_{-0.16}^{+0.17}$ \\
         Jitter $\sigma$ & ms$^{-1}$ & $1/\sigma$ (log-flat) & ${15.3}_{-1.2}^{+1.4}$ & ${4.1}_{-1.2}^{+1.1}$ \\
         $\sqrt{e}\sin\omega$ & & $\mathcal{U}(-1,1)$ & $-0.5457\pm0.0046$ & $0.5194\pm0.0088$ \\
         $\sqrt{e}\cos\omega$ & & $\mathcal{U}(-1,1)$ & ${-0.2182}_{-0.0081}^{+0.0078}$ & $0.5725\pm0.0067$ \\
         PA of the ascending node $\Omega$ & $\degree$ & $\mathcal{U}(-180,180)$ &$346.39_{-0.85}^{+0.87}$ & ${97.37}_{-0.36}^{+0.37}$ \\
         \hline
         & & Derived parameters \\
         \hline
         Orbital Period $P$ & years & & $21.59\pm0.05$ & ${50.9}_{-1.5}^{+1.7}$ \\
         Eccentricity $e$ & & & $0.346\pm0.004$ &  ${0.598}_{-0.008}^{+0.009}$ \\
         Argument of periastron $\omega$ & $\degree$ & & $248.20_{-0.86}^{+0.84}$ & $42.22\pm0.72$ \\
         Time of Periastron $T_0$ & JD & & $2458703\pm12$ & ${2473403}_{-563}^{+622}$ \\
         Semimajor axis & mas & & $206.5^{+3.6}_{-3.5}$ & ${317.5}_{-3.6}^{+3.9}$ \\
         Mass ratio $q$ & $M_{\rm{comp}}/M_{\rm{host}}$ & & $0.083^{+0.0015}_{-0.0014}$ & $0.079\pm0.0012$ \\
         Total mass $M_{\rm{total}}$ & $M_{\odot}$ & & $0.96^{+0.05}_{-0.04}$ & $1.05\pm0.03$ \\
         \hline \\
    \end{tabular}
    \label{tab:orbital_parameters}
\end{table*}

\subsection{HD~206505 (HIP~107665)} \label{sec:HD206505}

The star HD~206505 was monitored with the CORALIE survey between October 2001 and July 2023, covering 22 years of observations with 91 radial velocity measurements in total, as shown in Fig.~\ref{fig:HD206505_orbit}. Using the radial velocity measurements and the astrometric information from the HGCA, we were able to predict the relative astrometry of the companion as shown in Fig.~\ref{fig:astrometric_prediction}.

The RVs and astrometric acceleration joint analysis were previously presented in \citet{2023arXiv230316717B}. Here, we present the first direct detection of HD~206505~B and updated orbital parameters that incorporate new relative astrometry from these observations along with the RVs and HGCA data as presented in \citet{2023arXiv230316717B}.

We directly imaged HD~206505~B with VLT/SPHERE in the $H2$ and $H3$ bands on 2019-08-06 as part of program 0103.C-0199(A) (PI: Rickman) and with a total integration time of 8192~seconds. Additional follow-up imaging was performed on 2021-07-01 in the $K1$ and $K2$ bands with VLT/SPHERE as part of program 105.20SZ.001 (PI: Rickman) and with a total integration time of 6144~seconds. The resulting images are shown in Fig.~\ref{fig:images}.

Even though the orbital phase is not fully covered by the radial velocity measurements of HD~206505, there are ample measurements from when HD~206505~B passed through periastron, where the radial velocity is at a maximum (see Fig.~\ref{fig:HD206505_orbit}), which provides a strong constraint on the eccentricity and a relatively good constraint on the orbital period. Using the radial velocities as well as the astrometry from the HGCA and the relative astrometry from imaging, we performed an orbit fit using \texttt{orvara} as described in Section~\ref{sec:HD112863}. As for the case of HD~112863, we used a parallel-tempered MCMC with 15 temperatures; for each temperature, we used 100 walkers with 40,000 steps per walker thinned by a factor of 50. We used a log-flat prior on the host star mass in order to also measure the mass dynamically.

From this orbital fit, we determined an orbital period of ${50.9}_{-1.5}^{+1.7}$~years which is in agreement with the orbital period of $51.61\pm0.03$~years from \cite{2023arXiv230316717B}. We measured the dynamical mass of the primary to be $M_{\rm{host}}=0.97\pm0.03~M_{\odot}$, which is in agreement with \cite{2023arXiv230316717B} of $0.88\pm0.06~M_{\odot}$ derived using the stellar SED. The dynamical mass is also in agreement with the isochronal mass measured in this paper of $0.93\pm0.02~M_{\odot}$. We measured the dynamical mass of the companion to be $M_{\rm{comp}}=79.8\pm1.8~M_{\rm{Jup}}$, which is also in agreement with the RV and the astrometrically derived mass from \cite{2023arXiv230316717B} of $75.60\pm3.30~M_{\rm{Jup}}$. As mentioned in the previous section, we did not impose a Gaussian prior on the primary star mass when performing the joint orbital analysis. Despite this fact, the additional relative astrometric information from direct imaging yielded a more precise dynamical mass on the companion than reported in \cite{2023arXiv230316717B}. For HD~206505, the gain in precision in the dynamical mass is greater than for HD~112863, as there is less orbital phase coverage of HD~206505 from the RVs alone, and therefore the relative astrometry provides more of a constraint.

The resulting orbital fits are shown in Fig.~\ref{fig:HD206505_orbit}, and the full orbital parameters are listed in Table~\ref{tab:orbital_parameters}. The posteriors from the MCMC are shown in Fig.~\ref{fig:HD206505_corner}. The resulting relative astrometry in terms of projected angular separation and position angle are shown in Fig.~\ref{fig:HD206505_relsep_pa}. 

\begin{figure}[h]
    \centering
    \includegraphics[width=0.5\textwidth]{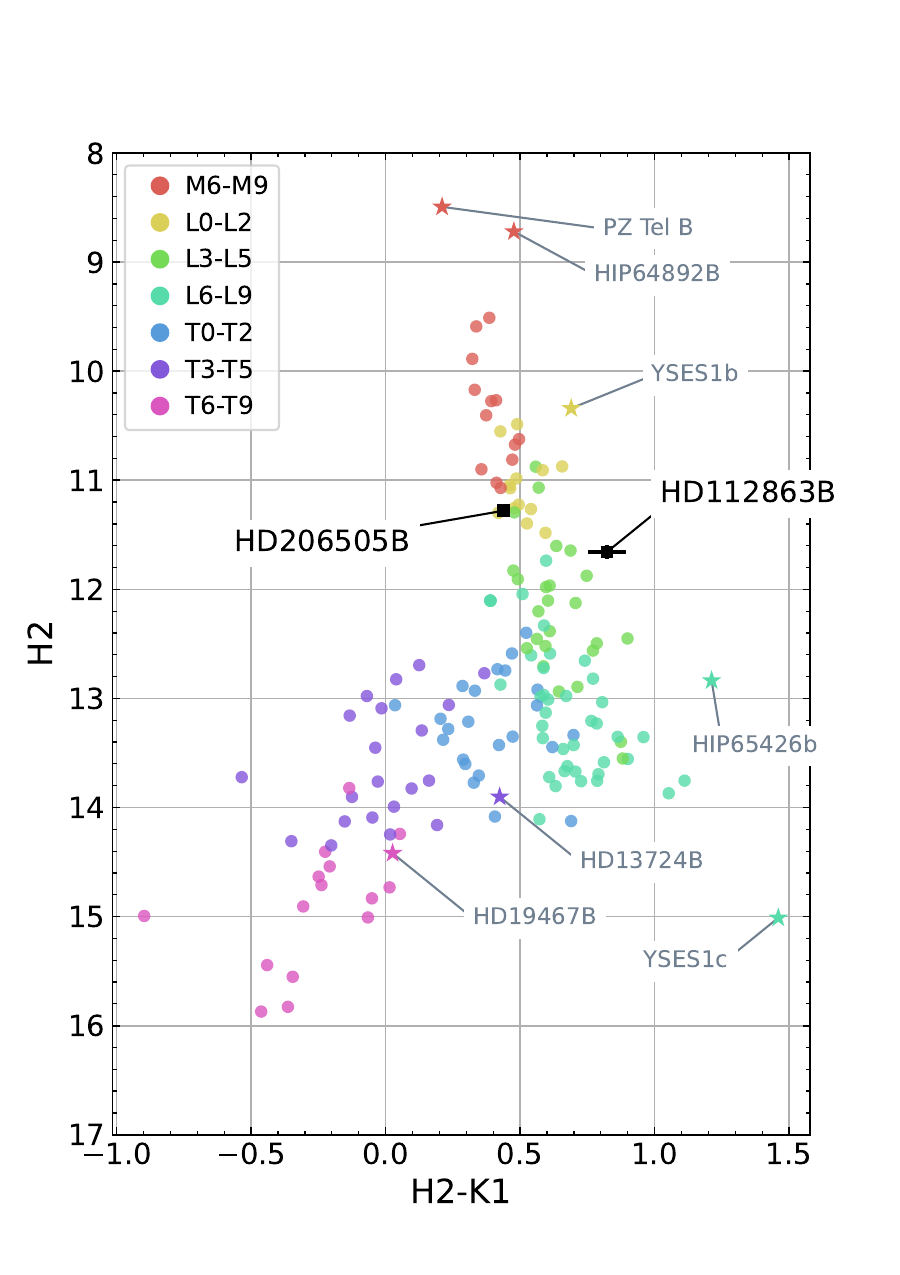}
    \caption{Color-magnitude diagram showing HD~112863~B and HD~206505~B (black squares) in comparison to the population of field brown dwarfs (circle symbols). We also include some notable substellar companions (star symbols). The field brown dwarfs are color-coded by spectral classification.}
    \label{fig:cmd}
\end{figure}

\section{Discussion} \label{sec:CMD}

We calculated the absolute flux of the companions by integrating a BT-NextGen model spectra \citep{2012RSPTA.370.2765A} of the target stars based on the stellar parameters given in Table~\ref{tab:stellar_params} through the filters and then multiplying the contrast values in Table~\ref{tab:astro_photo} while also accounting for the SPHERE filter transmission curves as described in Section~\ref{sec:imaging}. These model spectra were scaled by the distances and radii of the target stars (Table~\ref{tab:stellar_params}). 

The absolute flux values were used to generate the color-magnitude diagram (CMD) shown in Fig.~\ref{fig:cmd}. In the figure, we also show a selection of field brown dwarfs and highlight some notable substellar companions that have previously been imaged with SPHERE. The field brown dwarfs shown in Fig.~\ref{fig:cmd} are from the Brown Dwarf Spectral Survey \citep{2003ApJ...596..561M, 2007ApJ...658.1217M}, the L \& T Dwarf Archive \citep{2004AJ....127.3516G,2004AJ....127.3553K,2006AJ....131.2722C}, and the IRTF Spectral Library \citep{2005ApJ...623.1115C, 2009ApJS..185..289R}, with updated distances. We included only those objects with parallactic distance measurements. Distances are from the Gaia data releases eDR3 or DR2 \citep{2016A&A...595A...1G,2018A&A...616A...1G,2021A&A...649A...1G} when available and from brown dwarf parallax studies otherwise \citep{2012ApJS..201...19D, 2012ApJ...752...56F, 2013MNRAS.433.2054S, 2014ApJ...796...39T, 2016ApJ...833...96L, 2017ApJS..231...15D, 2018MNRAS.481.3548S, 2020AJ....159..257B}. Photometry for the highlighted substellar companions of interest is taken from \citet{2016A&A...587A..56M, 2017A&A...605L...9C, 2018A&A...615A.160C, 2020A&A...639A..47M, 2020A&A...635A.203R, 2020MNRAS.492..431B, 2020ApJ...898L..16B}.

The companions HD~206505~B and HD~112863~B are ``twin" objects, and both objects are consistent with those of early-L field brown dwarfs, which are well above the L-T transition. Both companions are fainter than any of the M-type objects plotted in Fig.~\ref{fig:cmd}, hinting that they are likely to be of substellar nature. Both brown dwarfs have very similar $K$ band absolute magnitudes (Fig.~\ref{fig:cmd}, see also Table~\ref{tab:astro_photo}); this is consistent with the fact that both companions have very similar dynamical masses, ages, and host star metallicities. HD~206505~B is toward the blue end of the distribution of field brown dwarfs, which agrees with predictions for a relatively massive, relatively old brown dwarf that has undergone significant cooling and contraction since formation. HD~112863~B is somewhat redder than expected for a field-age brown dwarf, which could potentially be due to the object having a lower surface gravity or hosting a dusty circumplanetary disk. Further investigations into the spectroscopic properties of these objects are being explored in detail in a follow-up paper (Ceva et al. in prep.) that will include verifying the red nature of HD~112863~B with wider wavelength coverage and higher resolution from SPHERE/IFS data.

\section{Summary} \label{sec:conclusion}

We report the direct detection of two new benchmark brown dwarfs respectively orbiting HD~112863 and HD~206505. Both companions have dynamical masses close to the stellar-substellar boundary. The dynamical masses of HD~112863~B and HD~206505~B are $77.1^{+2.9}_{-2.8}~M_{\textrm{Jup}}$ and $79.8\pm1.8~M_{\textrm{Jup}}$, respectively. 

These masses were calculated through orbit fitting with \texttt{orvara} by combining the relative astrometry determined from direct imaging with VLT/SPHERE with radial velocity measurements from CORALIE as well as with astrometry from \emph{Hipparcos} and \emph{Gaia}. We measured the precise model-independent masses of these brown dwarf companions, which are vital benchmark objects to probe the stellar-substellar boundary. These objects can be used to empirically validate mass-luminosity-age relations of substellar objects that are degenerate in nature and contain a number of underlying assumptions, and they join a small but growing list of known benchmark substellar companions \cite[e.g.,][]{2018A&A...614A..16C,2019A&A...631A.107P,2020A&A...635A.203R,2020A&A...639A..47M,2022MNRAS.513.5588B,2022AJ....163...50F,2023AJ....165...39F} that serve as key calibrators of brown dwarf evolutionary models. Furthermore, the result of these direct detections validates the strategy for direct imaging of exoplanets and brown dwarfs by using precursor information such as radial velocities and/or astrometry to select candidates based on the potential for direct detectability, increasing the detection efficiency of such objects, as demonstrated in Fig.~\ref{fig:astrometric_prediction}.

The dynamical masses of the brown dwarf companions are both in agreement with the recently published values from \cite{2023arXiv230316717B} that were calculated from the radial velocity measurements and proper motions. The relative astrometry derived from the direct detections in this paper provides further constraints on the orbital parameters both in terms of measuring a dynamical mass on each of the primary stars and determining a higher level of precision on the companion dynamical masses.

As we did not use any informed priors on the host star masses for the orbital fits, we were able to measure the dynamical masses of the host stars. This approach is unlike that of \cite{2023arXiv230316717B}, who imposed a Gaussian prior on the host stars in order to perform the orbital fit. The host star dynamical masses we report are in agreement with the masses determined by \cite{2023arXiv230316717B} using the stellar SEDs as well as the stellar masses measured in this paper using stellar isochrones, as described in Section~\ref{sec:stellar_params} and shown in Table~\ref{tab:stellar_params}. We were also able to measure the dynamical masses of the companions to a higher level of precision, with an improvement from a $\sim4.4\%$ error reported in \cite{2023arXiv230316717B} for both HD~112863~B and HD~206505~B to errors of $3.6\%$ and $2.2\%$ for HD~112863~B and HD~206505~B, respectively.

Using the $H$ and $K$ band photometry from VLT/SPHERE, we determine that both HD~112863~B and HD~206505~B are early-mid L-types, as shown in Fig.~\ref{fig:cmd}. More extensive follow-up of the spectroscopic properties of these two new benchmark brown dwarfs will be presented in Ceva et al. (in prep.) in order to explore the nature of these objects in more detail and to test against brown dwarf evolutionary models.

\begin{acknowledgements}
This work has been carried out within the framework of the National Centre for Competence in Research PlanetS supported by the Swiss National Science Foundation. The authors acknowledge the financial support of the SNSF. This publication makes use of the The Data \& Analysis Center for Exoplanets (DACE), which is a facility based at the University of Geneva (CH) dedicated to extrasolar planets data visualization, exchange and analysis. DACE is a platform of the Swiss National Centre of Competence in Research (NCCR) PlanetS, federating the Swiss expertise in Exoplanet research. The DACE platform is available at \url{https://dace.unige.ch}. This work has made use of data from the European Space Agency (ESA) mission Gaia (\url{https://www.cosmos.esa.int/gaia}), processed by the Gaia Data Processing and Analysis Consortium (DPAC, \url{https://www.cosmos.esa.int/web/gaia/dpac/consortium}). Funding for the DPAC has been provided by national institutions, in particular the institutions participating in the Gaia Multilateral Agreement. This research made use of the SIMBAD database and the VizieR Catalogue access tool, both operated at the CDS, Strasbourg, France. The original descriptions of the SIMBAD and VizieR services were published in \citet{2000A&AS..143....9W} and \citet{2000A&AS..143...23O}. This research has made use of NASA’s Astrophysics Data System Bibliographic Services.
\end{acknowledgements}

%
\bibliography{bib.bib} 
%

\begin{appendix}

\section{DACE links} \label{appendix_a}

The radial velocity measurements and the additional data products discussed in this paper are available in electronic form on the DACE web platform for each individual target:

\begin{itemize}
\item HD~112863: \url{https://dace.unige.ch/radialVelocities/?pattern=HD112863}
\item HD~206505: \url{https://dace.unige.ch/radialVelocities/?pattern=HD206505}
\end{itemize}

\section{Posterior distributions of the orbital fits} \label{appendix_posterior_distributions}

Here we show the corner plots for the posterior distributions of the orbital fits for each system fitted using \texttt{orvara}. For each system, the posterior distributions for the primary stellar mass ($M_{\rm{pri}}$), the companion mass ($M_{\rm{sec}}$), the semimajor axis ($a$), the eccentricity ($e$), and the orbital inclination ($i$) are shown.

\begin{figure*}
    \centering
    \includegraphics[width=\textwidth]{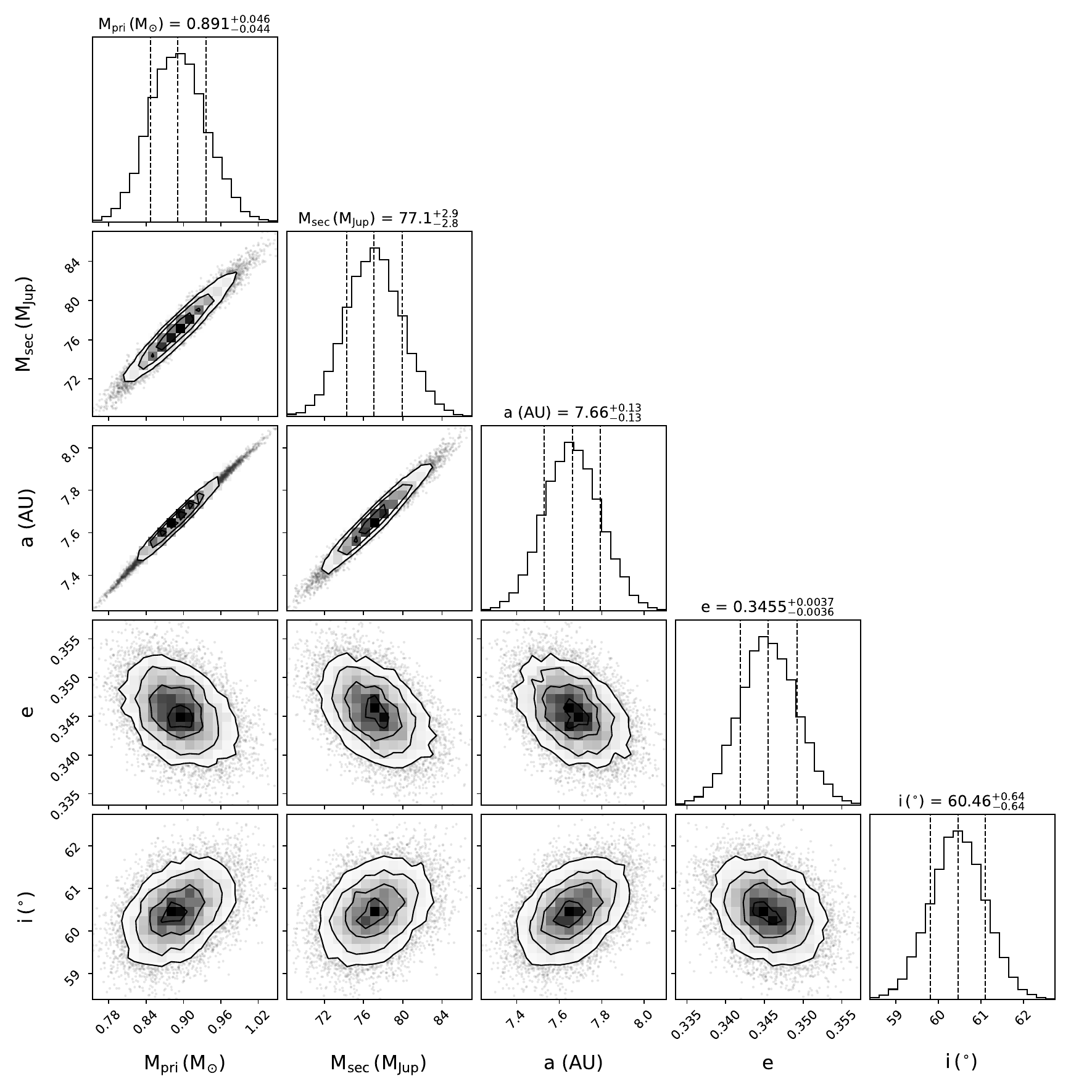}
    \caption{Marginalized 1D and 2D posterior distributions for selected orbital parameters of HD~112863~B corresponding to the fit of the RV, relative astrometry from direct imaging observations, and absolute astrometry from \emph{Hipparcos} and Gaia with the use of \texttt{orvara}. Confidence intervals at 15.85\%, 50.0\%, 84.15\% are overplotted on the 1D posterior distributions; the median $\pm1\sigma$ values are given at the top of each 1D distribution. The 1, 2, and 3$\sigma$ contour levels are overplotted on the 2D posterior distribution.}
    \label{fig:HD112863_corner}
\end{figure*}

\begin{figure*}
    \centering
    \includegraphics[width=\textwidth]{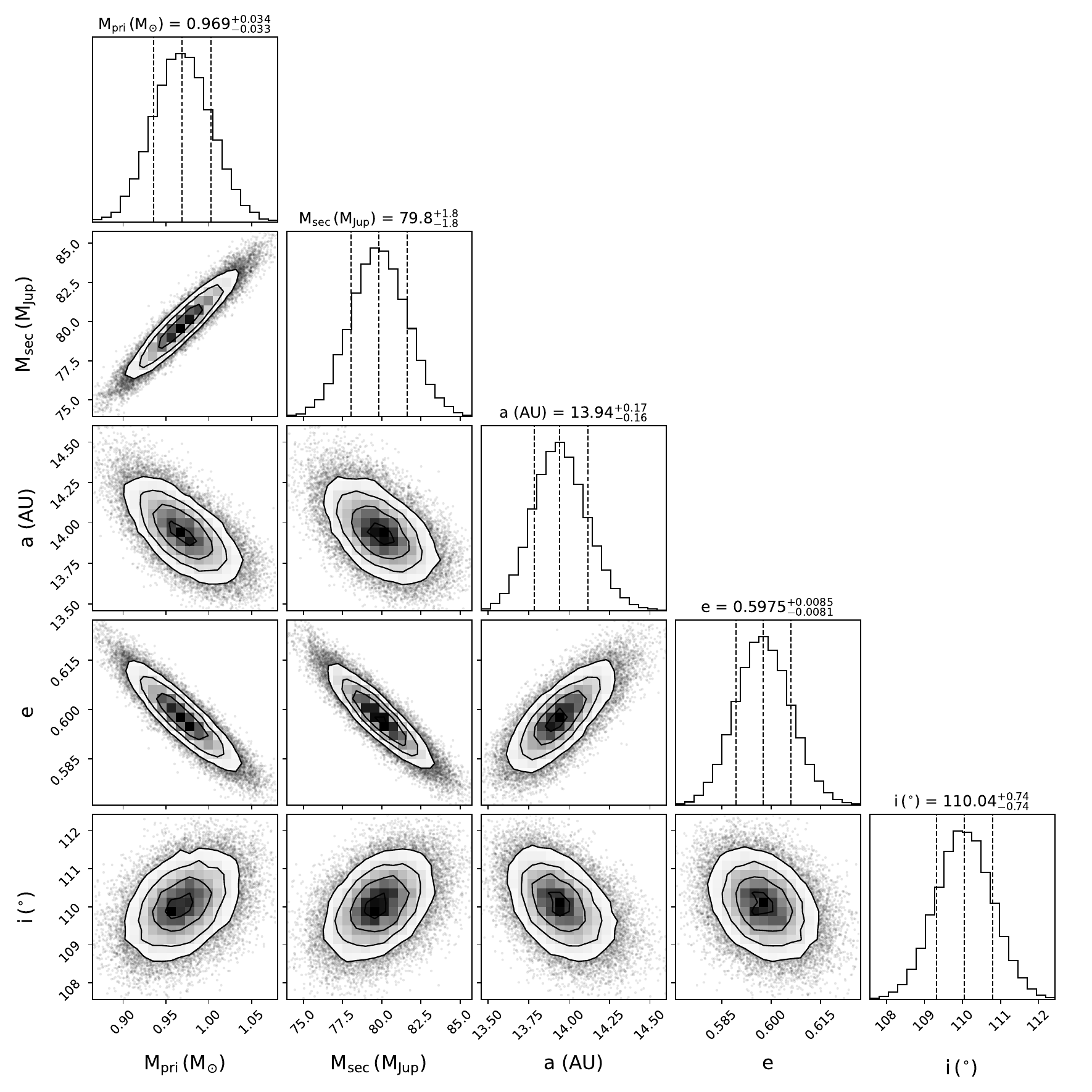}
    \caption{Marginalized 1D and 2D posterior distributions for selected orbital parameters of HD~206505~B corresponding to the fit of the RV, relative astrometry from direct imaging observations, and absolute astrometry from \emph{Hipparcos} and Gaia with the use of \texttt{orvara}. Confidence intervals at 15.85\%, 50.0\%, 84.15\% are overplotted on the 1D posterior distributions; the median $\pm1\sigma$ values are given at the top of each 1D distribution. The 1, 2, and 3$\sigma$ contour levels are overplotted on the 2D posterior distribution.}
    \label{fig:HD206505_corner}
\end{figure*}

\section{Relative astrometry} \label{appendix_relative_astrometry}

\begin{figure*}
    \centering
    \includegraphics[width=0.49\textwidth]{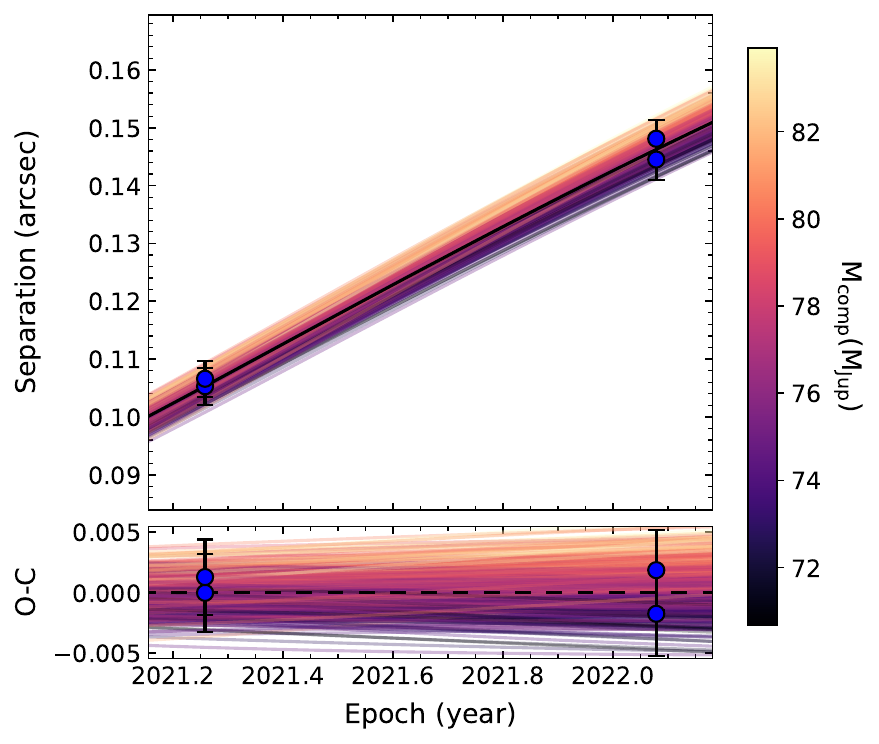}
    \includegraphics[width=0.48\textwidth]{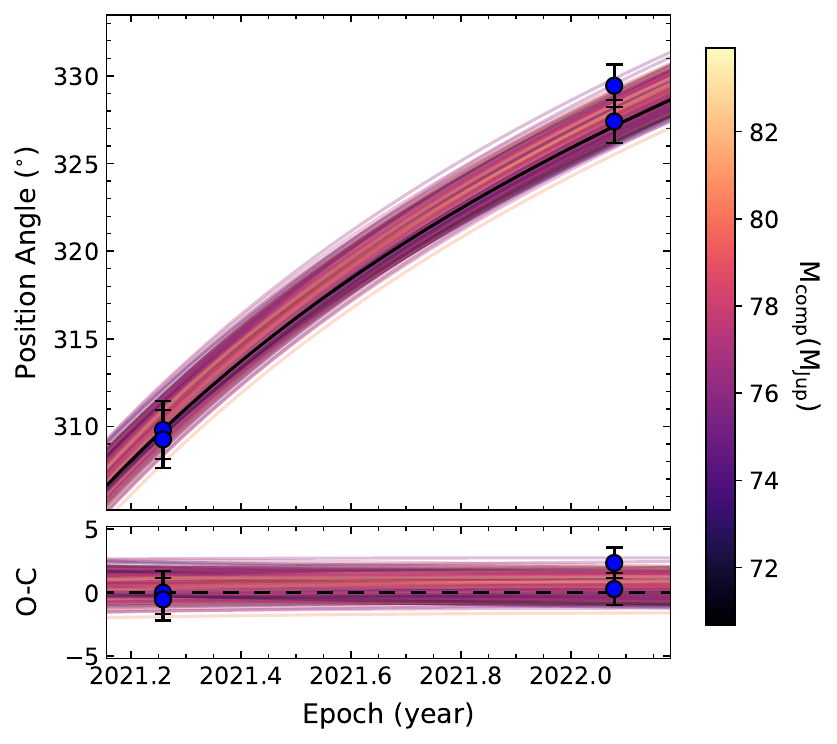}
    \caption{Relative separation (left) and position angle (right) of HD~112863~B as a function of time across two epochs of observations corresponding to the values in Table~\ref{tab:astro_photo}. The thick black line represents the highest likelihood orbit. The thin colored lines are 500 orbits drawn randomly from the posterior distribution. Darker purple indicates a lower companion mass, and light yellow represents a higher companion mass. The residuals of the proper motions are shown in the bottom panels.}
    \label{fig:HD112863_relsep_pa}
\end{figure*}

\begin{figure*}
    \centering
    \includegraphics[width=0.50\textwidth]{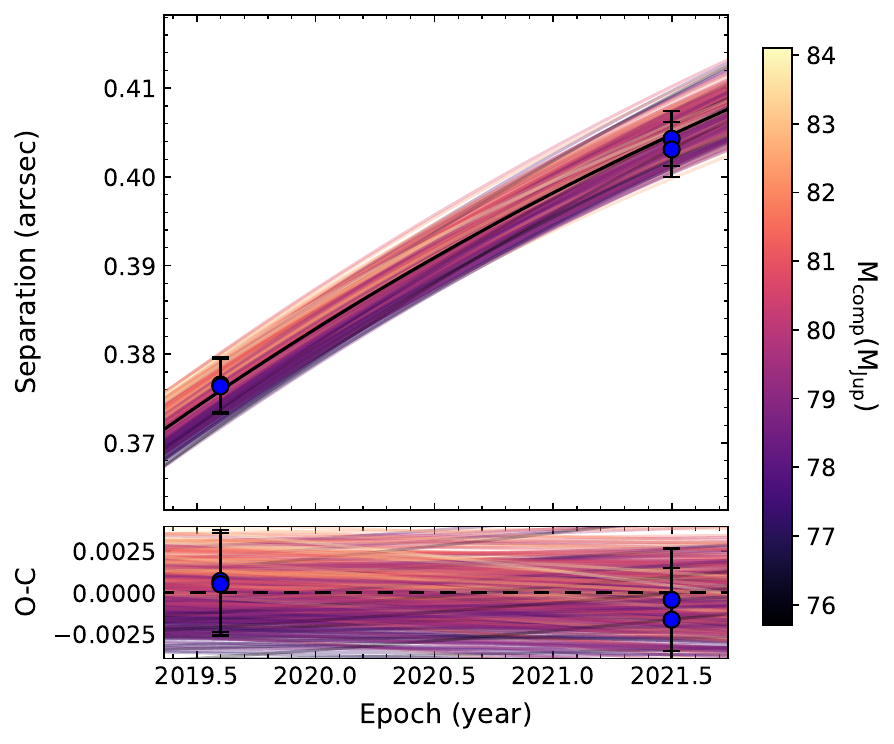}
    \includegraphics[width=0.48\textwidth]{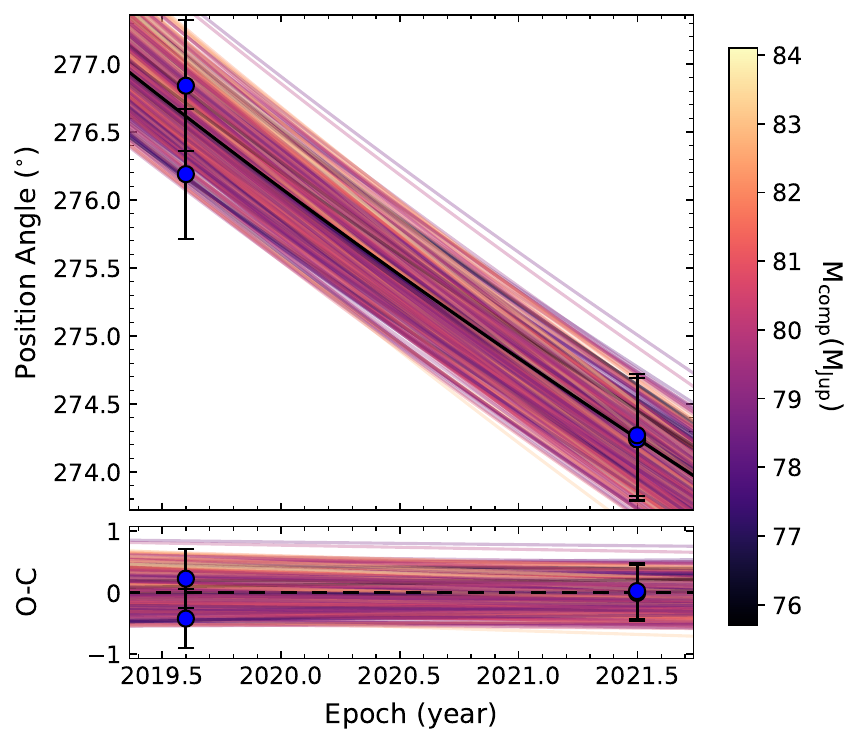}
    \caption{Same as Fig.~\ref{fig:HD112863_relsep_pa} but for HD~206505~B.}
    \label{fig:HD206505_relsep_pa}
\end{figure*}

Here we show the plots of the relative astrometry in projected angular separation (arcsec) and position angle (deg) across the two epochs of data for both companions. The fits to the relative astrometry shown here come from the orbital fitting using \texttt{orvara}. The plots for HD~112863~B are shown in Fig.~\ref{fig:HD112863_relsep_pa}, and the plots for HD~206505~B are shown in Fig.~\ref{fig:HD206505_relsep_pa}, with the corresponding residuals shown across the bottom panels of each plot.

\section{Astrometric predictions} \label{appendix_astrometric_predictions}

\begin{figure*}
    \centering
    \includegraphics[width=0.498\textwidth]{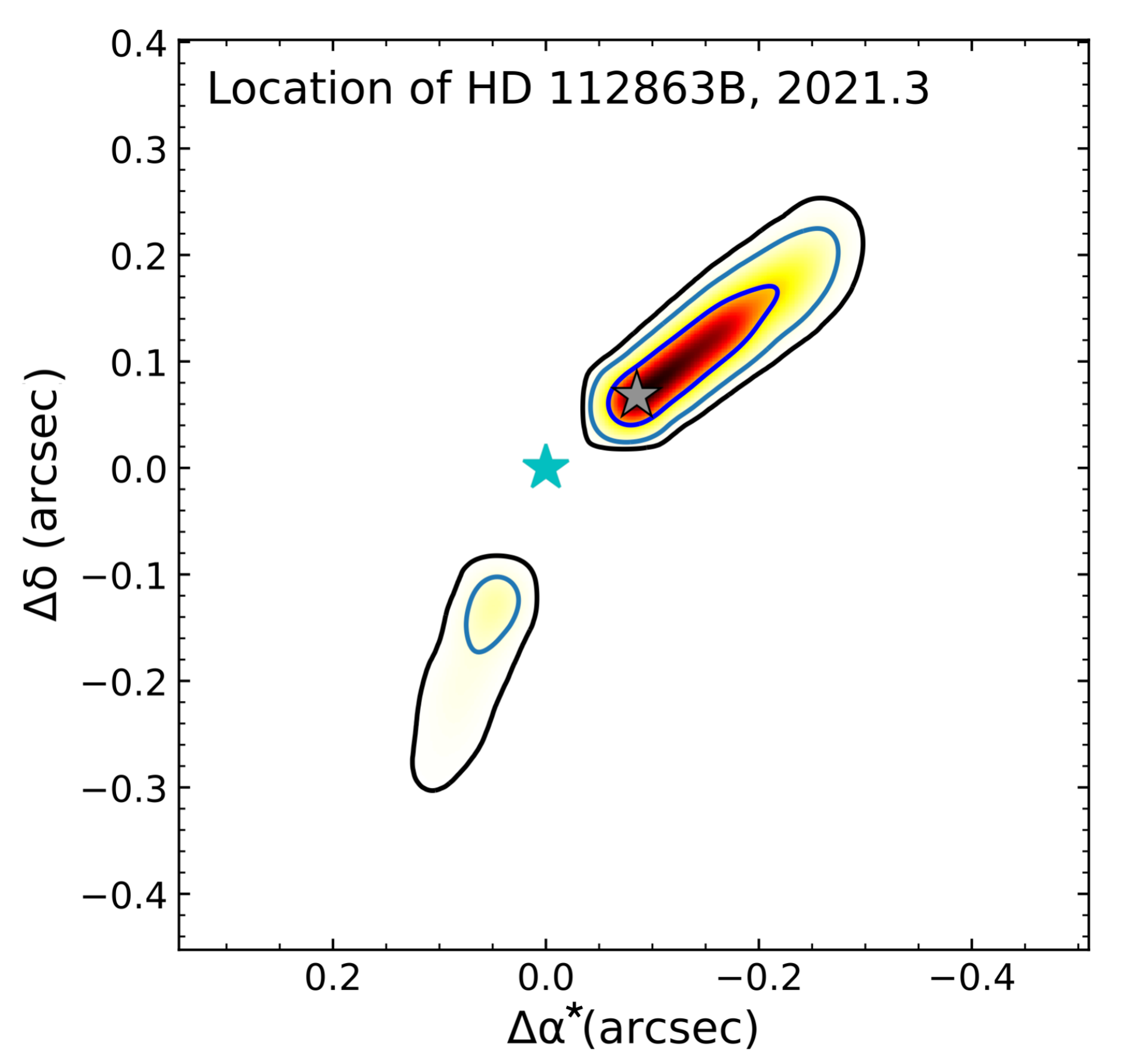}
    \includegraphics[width=0.475\textwidth]{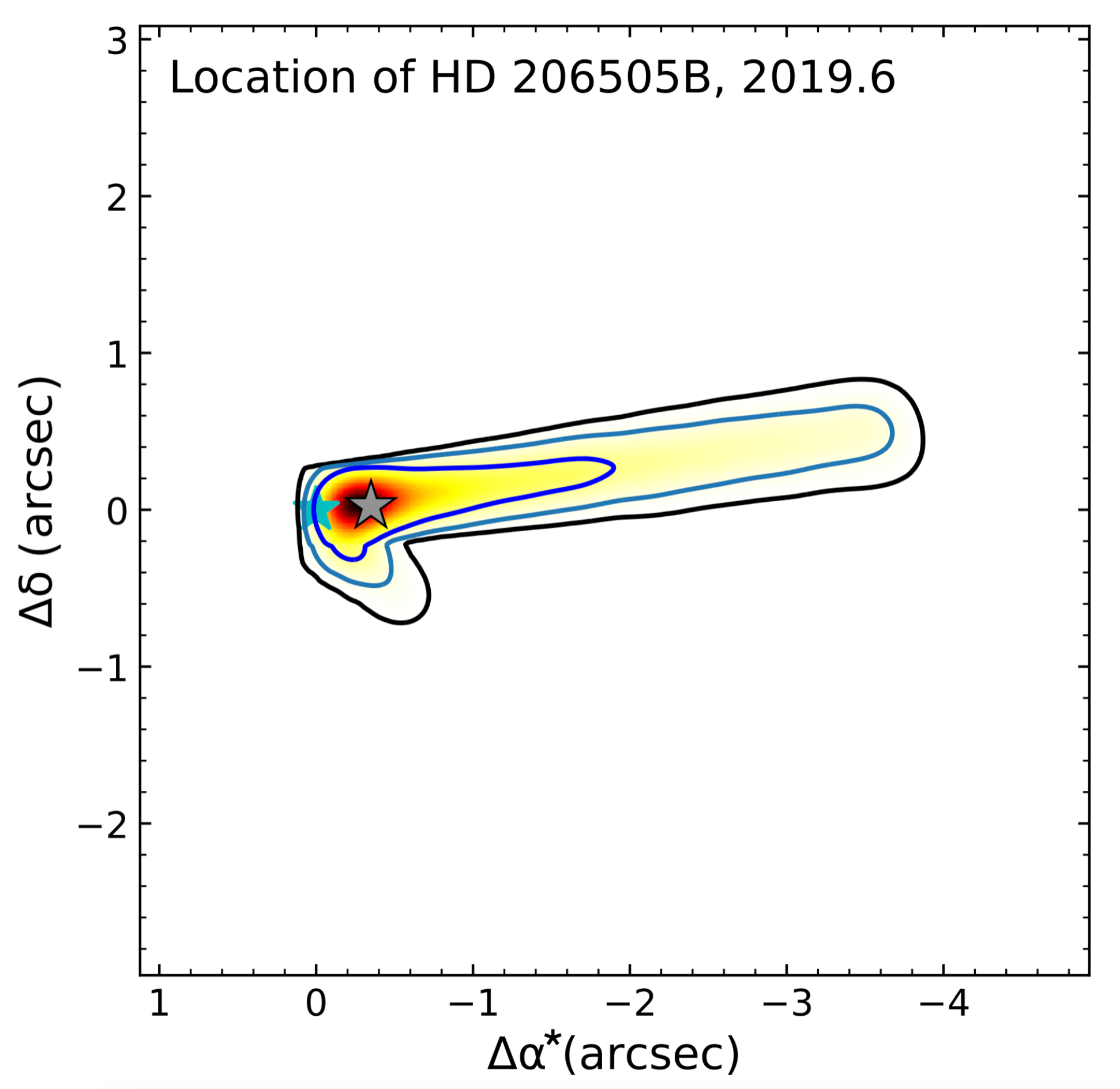}
    \caption{Predicted relative astrometric positions for HD~112863~B (\textbf{left}) and HD~206505~B (\textbf{right}) from orbital fits using the radial velocity and HGCA data relative to their host stars in right ascension ($\Delta \alpha^{*} = \Delta \alpha \cos \delta$) and declination ($\Delta \delta$). The blue star represents the primary star, and the gray star shows the position of the detected companion relative to the host star in the epoch of the first direct detection of each companion (2021.3 and 2019.6 for HD~112863~B and HD~206505~B respectively.) The contour lines represent the positions to 1, 2, and $3~\sigma$ predicted positions of each companion calculated for the time of the first direct observations of each target.}
    \label{fig:astrometric_prediction}
\end{figure*}

Using the radial velocities and the astrometric accelerations from the HGCA, we perform orbit fits to indicate the predicted relative astrometric position of both companions. In Fig.~\ref{fig:astrometric_prediction} we show the predicted positions for both HD~112863 and HD~206505 relative to their host stars in right ascension ($\Delta \alpha^{*} = \Delta \alpha \cos \delta$) and declination ($\Delta \delta$) using the \texttt{orvara} orbit-fitting package, with the measured companion positions shown by the overplotted gray stars. Relative astrometric positions indicate whether a companion can be directly detected with imaging given its projected separation between the host star and the companion itself. The 1, 2, and 3$\sigma$ contour levels of the predicted positions of each companion are plotted from a full orbit fit using just the radial velocities and the HGCA astrometry. The astrometric predictions are shown for the epoch in which each companion was then subsequently directly imaged with VLT/SPHERE to visually compare the prior predictions to the direct detections. The measured positions of both companions from direct imaging observations agree well with the predicted positions from the orbit fitting.

\section{Spectroscopic analysis} \label{appendix_RV}

For both HD~112863 and HD~206505, it was important to ensure that the radial velocity measurements were induced by the presence of the brown dwarf companions and not by stellar activity nor other unseen planetary companions. Therefore, we carried out additional analyses of the spectroscopic stellar activity and radial velocity data.

\subsection{Stellar activity indicators}

\begin{figure*}
    \centering
    \includegraphics[width=0.9\textwidth]{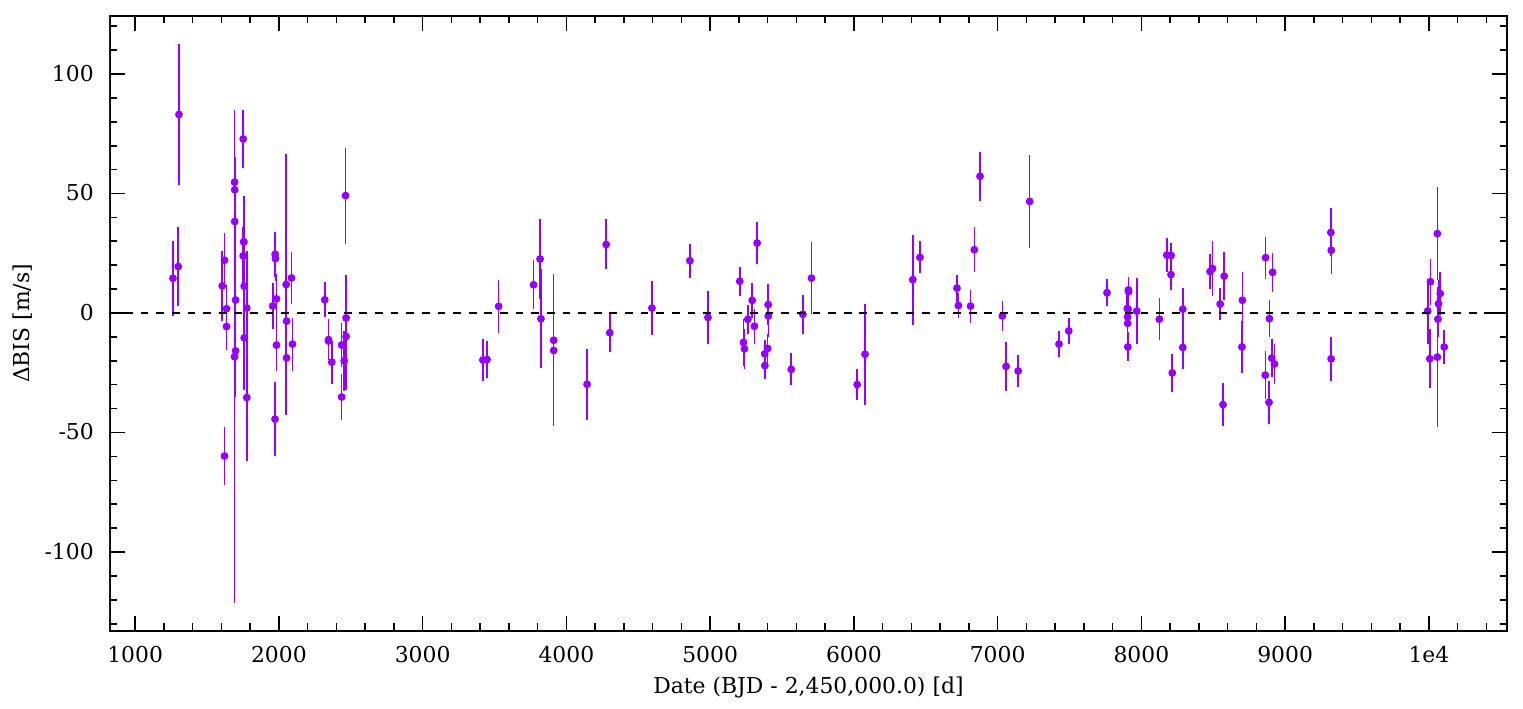}
    \includegraphics[width=0.9\textwidth]{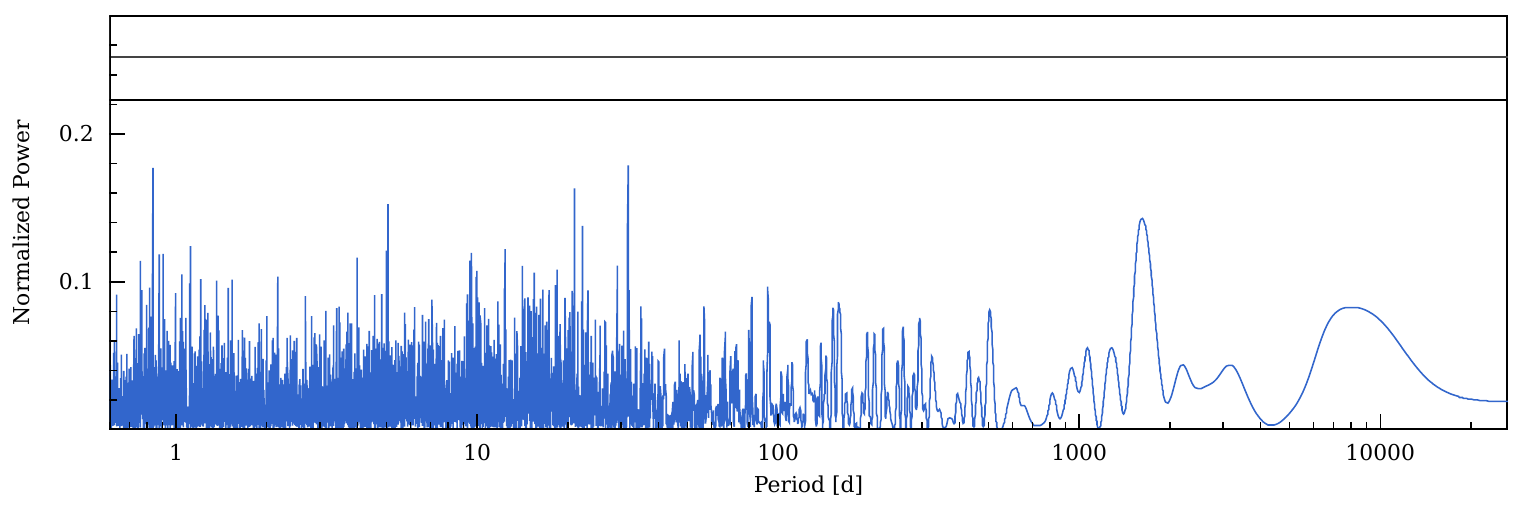}
    \caption{Spectroscopic analysis of HD~112863~A using the CCF bisector. \textbf{Top:} Measured CCF bisector of the host star HD~112863 as a function of time. \textbf{Bottom:} Corresponding Lomb-Scargle periodogram. The two black horizontal lines show the 1\% (bottom) and the 0.1\% (top) FAPs and demonstrate that there are no significant periodic signals in the CCF bisector indicator.}
    \label{fig:HD112863_CCF}
\end{figure*}

\begin{figure*}
    \centering
    \includegraphics[width=0.9\textwidth]{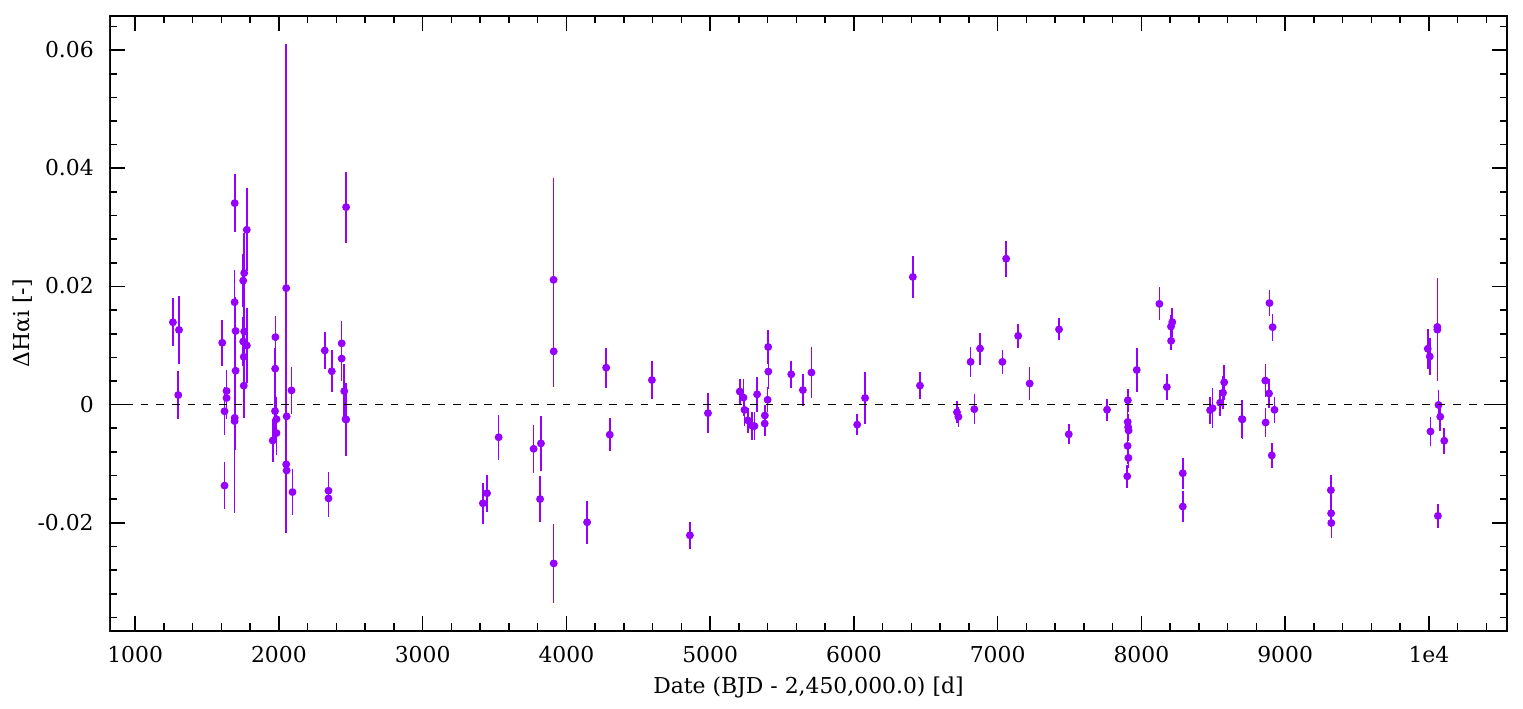}
    \includegraphics[width=0.9\textwidth]{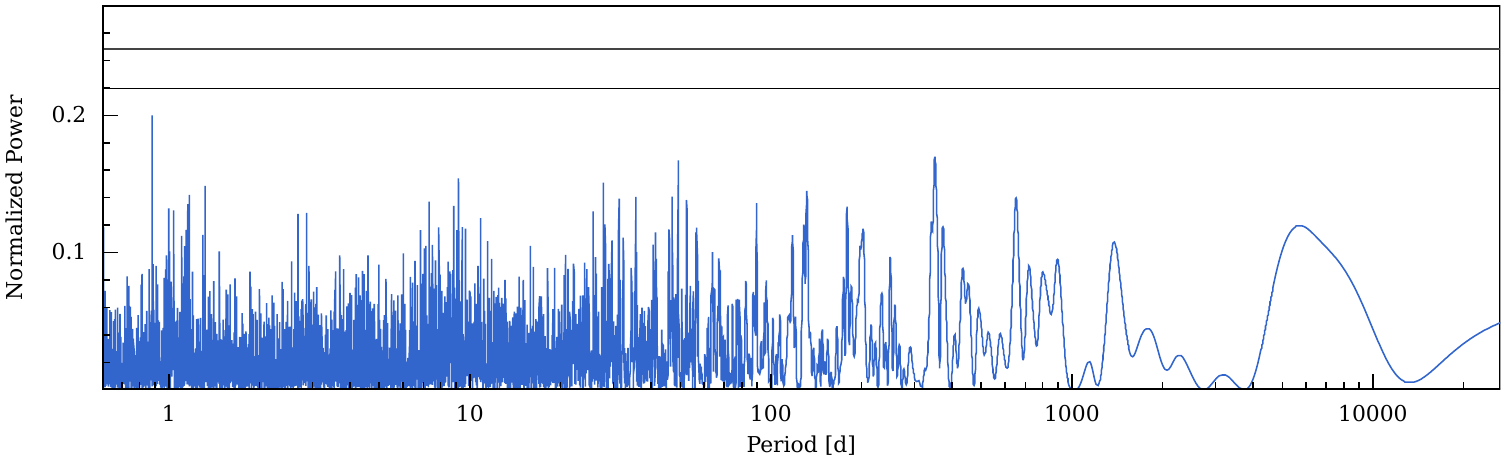}
    \caption{Spectroscopic analysis of HD~112863~A using the $H_{\alpha}$ chromospheric index. \textbf{Top:} Measured $H_{\alpha}$ chromospheric index of the host star HD~112863 as a function of time. \textbf{Bottom:} Corresponding Lomb-Scargle periodogram. The two black horizontal lines show the 1\% (bottom) and the 0.1\% (top) FAPs and demonstrate that there are no significant periodic signals in the $H_{\alpha}$ chromospheric indicator.}
    \label{fig:HD112863_Halpha}
\end{figure*}

\begin{figure*}
    \centering
    \includegraphics[width=0.9\textwidth]{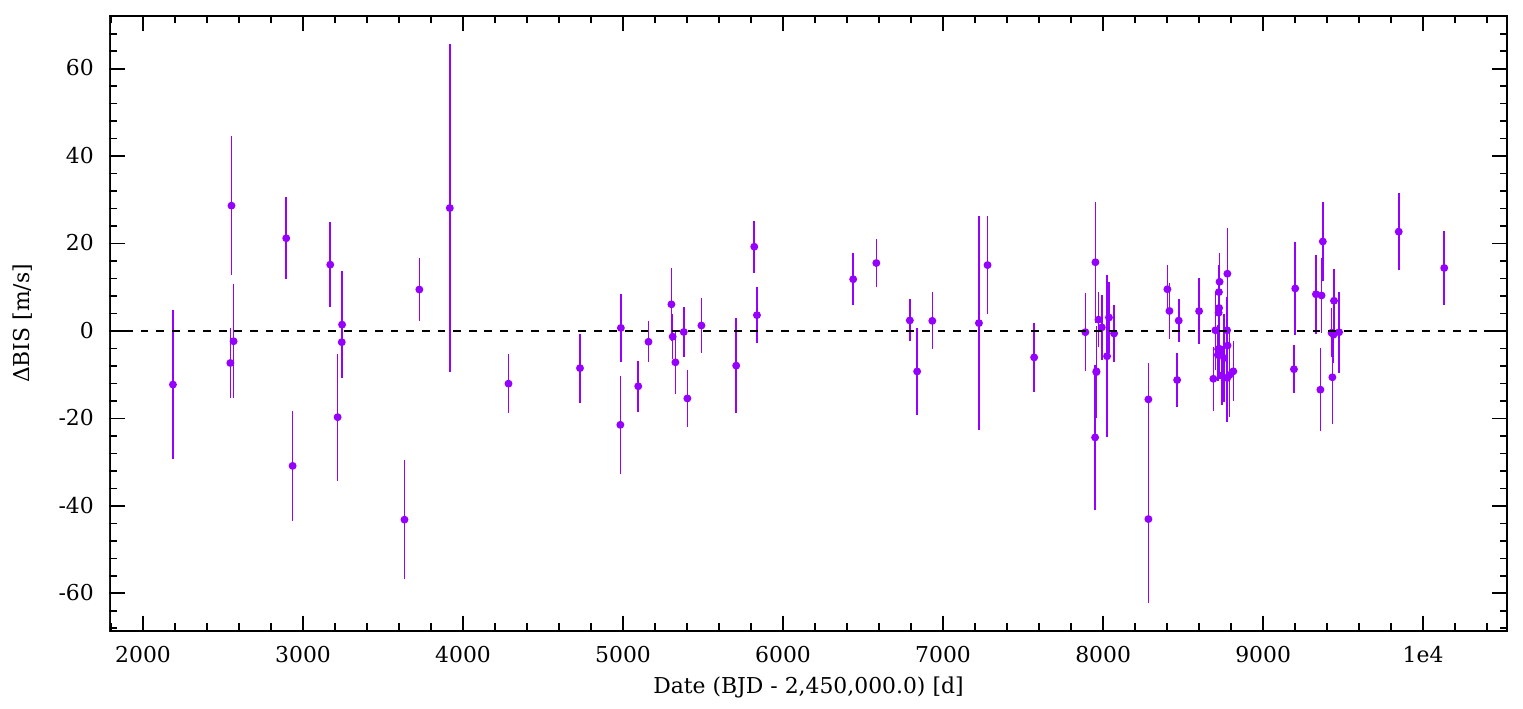}
    \includegraphics[width=0.9\textwidth]{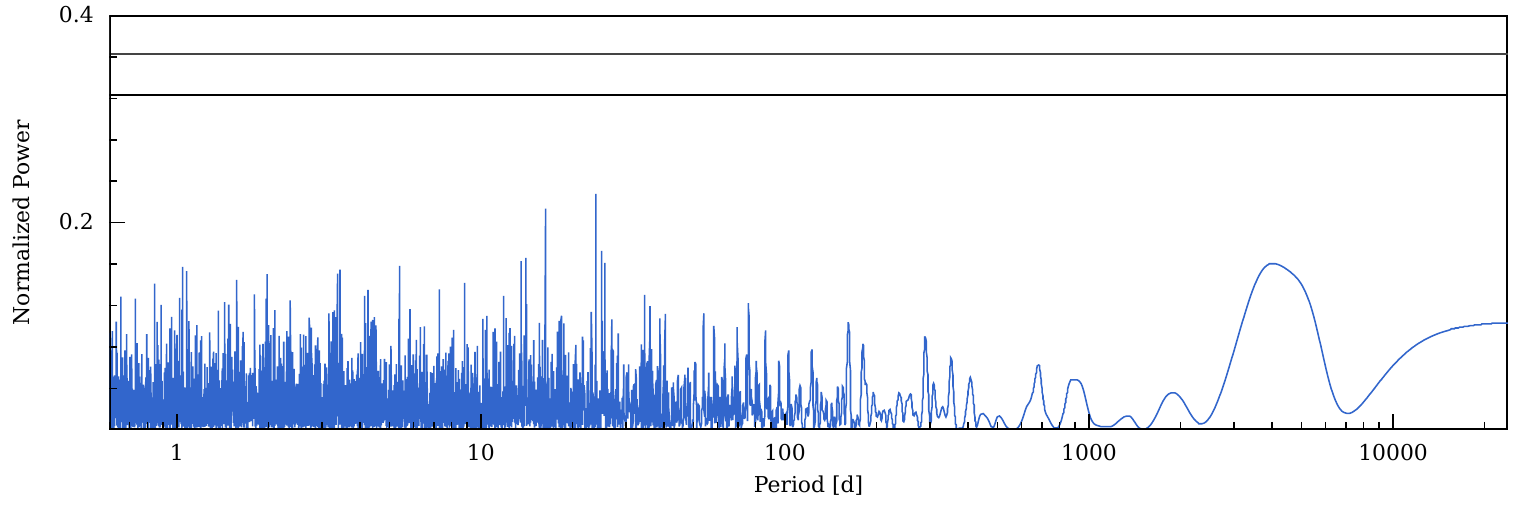}
    \caption{Same as Fig.~\ref{fig:HD112863_CCF} but for HD~206505.}
    \label{fig:HD206505_CCF}
\end{figure*}

\begin{figure*}
    \centering
    \includegraphics[width=0.9\textwidth]{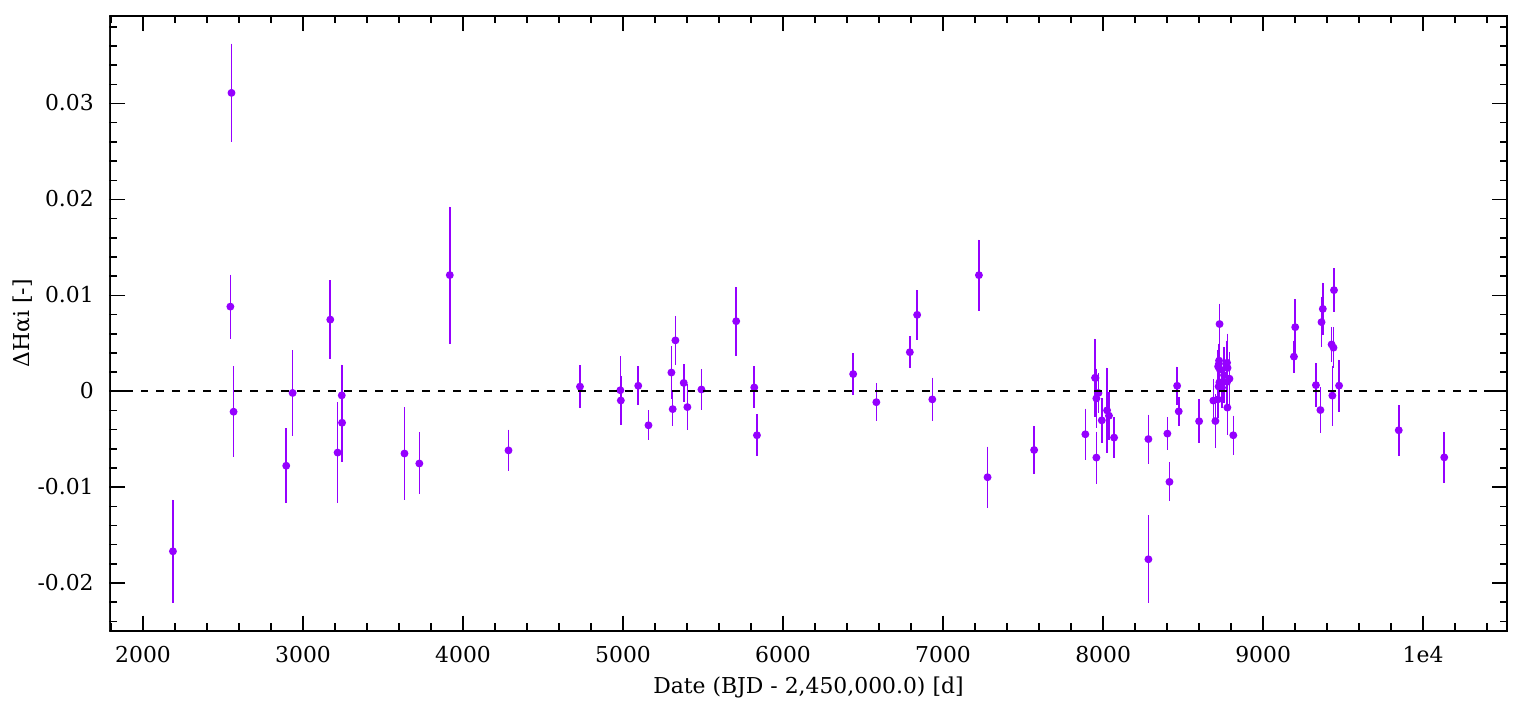}
    \includegraphics[width=0.9\textwidth]{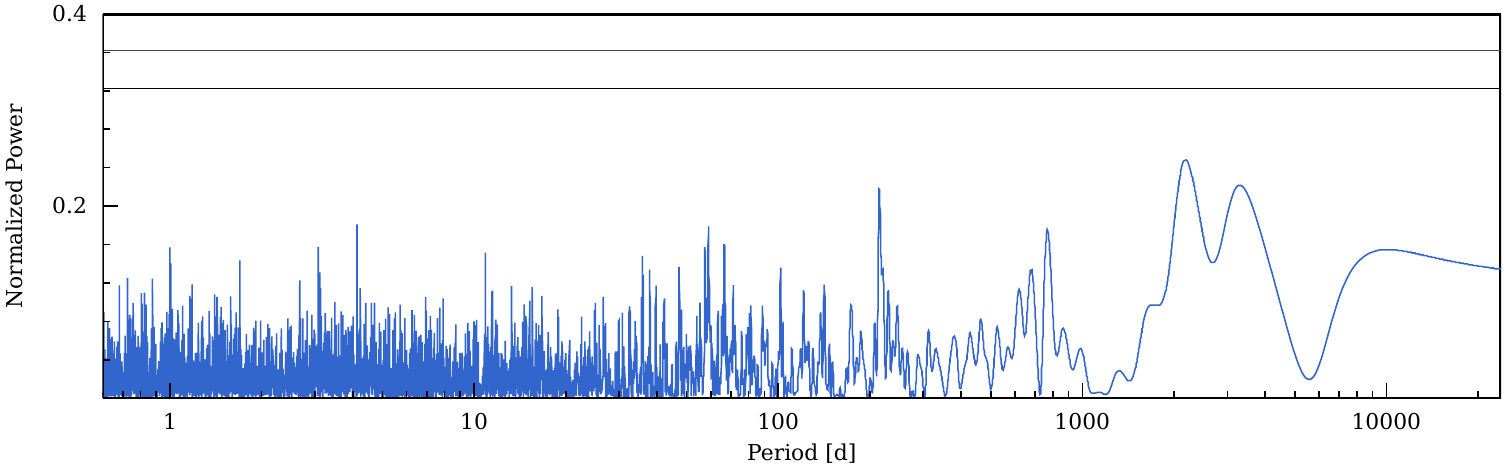}
    \caption{Same as Fig.~\ref{fig:HD112863_Halpha} but for HD~206505.}
    \label{fig:HD206505_Halpha}
\end{figure*}

To rule out any stellar activity of the host star that could mimic the expected radial velocity of the brown dwarf companions, we carried out an analysis of the CCF bisector and the $H_{\alpha}$ chromospheric activity indicator for both systems. In Figs.~\ref{fig:HD112863_CCF}, \ref{fig:HD112863_Halpha}, \ref{fig:HD206505_CCF}, and \ref{fig:HD206505_Halpha}, we show the measured stellar activity indicator as a function of time with the corresponding generalized Lomb-Scargle periodograms \citep{2009A&A...496..577Z} as well as the computed analytical false-alarm probabilities (FAPs), which were obtained following the methodology of \citet{2008MNRAS.385.1279B}. Any periodic signals with a FAP lower than 0.1\% are considered significant. We computed the 1\% and the 0.1\% FAP for both systems and did not see any significant signals in either the CCF bisector or the $H_{\alpha}$ chromospheric index. This supports the evidence that the long-term variation seen in the radial velocity data is due to acceleration from a companion rather than any activity effect of the host star.

\subsection{Radial velocity}

\begin{figure*}
    \centering
    \includegraphics[width=0.9\textwidth]{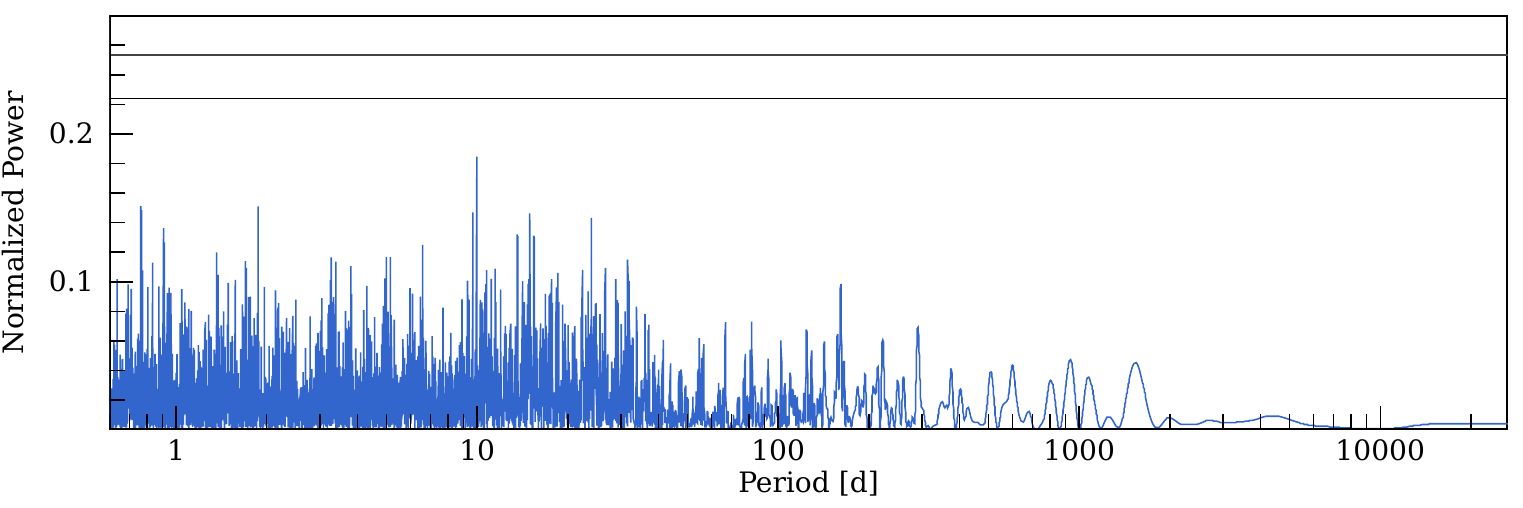}
    \includegraphics[width=0.9\textwidth]{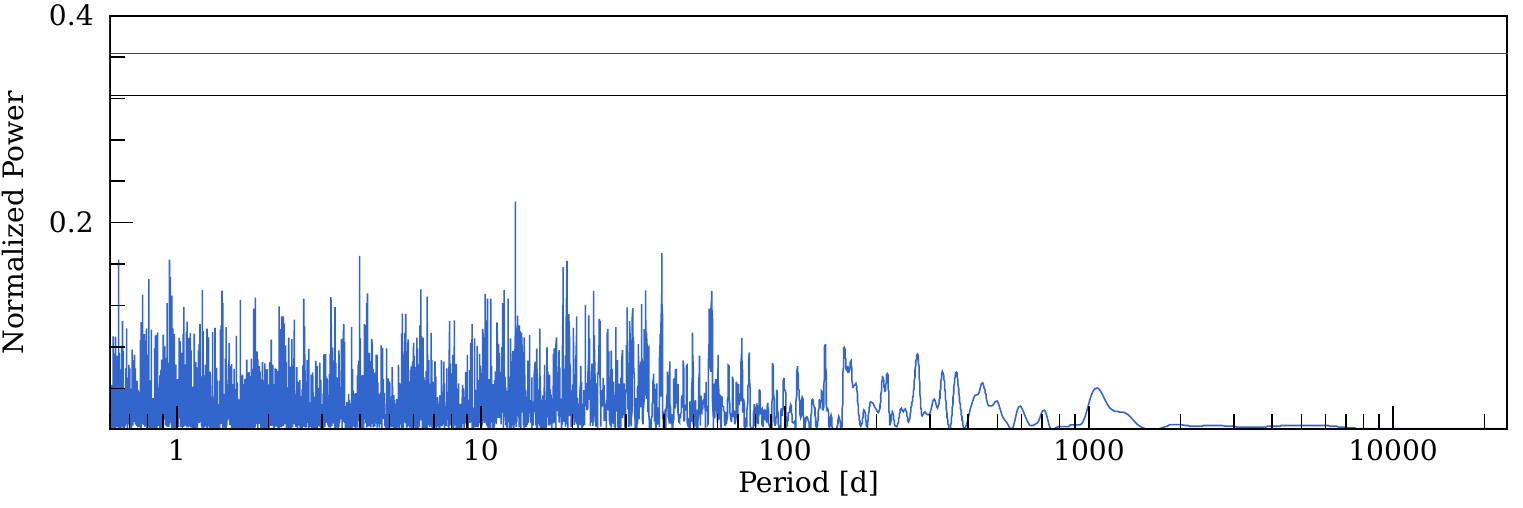}
    \caption{Periodograms of the radial velocity residuals for HD~112863 (\textbf{top}) and HD~206505 (\textbf{bottom}) after the brown dwarf companion radial velocity signal has been fitted as shown in Figs.~\ref{fig:HD112863_orbit} and \ref{fig:HD206505_orbit}. The black horizontal lines correspond to the 1\% (bottom) and 0.1\% (top) FAPs and demonstrate that there are no other significant radial velocity signals in the data, ruling out the possibility of additional companions in either system.}
    \label{fig:RV_periodograms}
\end{figure*}

We show the periodograms for the radial velocity residuals for both HD~112863 and HD~206505 in Fig.~\ref{fig:RV_periodograms}. These correspond to the residuals after the RVs induced by the brown dwarf companions have been fitted and removed, as shown in Figs.~\ref{fig:HD112863_orbit} and \ref{fig:HD206505_orbit}. As seen in Fig.~\ref{fig:RV_periodograms}, there are no significant RV signals in the residuals beyond a 1\% FAP, and therefore we concluded that there are no additional planetary companions in either system.

\end{appendix}

\end{document}